\PassOptionsToPackage{unicode}{hyperref}
\PassOptionsToPackage{hyphens}{url}
\PassOptionsToPackage{dvipsnames,svgnames,x11names}{xcolor}
\documentclass[12pt]{article}

\usepackage{varwidth}
\usepackage{threeparttable}
\usepackage{makecell}
\usepackage{amsfonts}
\usepackage{mathrsfs}
\usepackage{fancyhdr}
\usepackage{epsfig,morefloats}
\usepackage[round]{natbib}
\usepackage{booktabs,multirow,multicol,longtable}
\usepackage{amsthm}
\usepackage{amssymb}
\usepackage{lscape}
\usepackage{rotating}
\usepackage{caption}
\usepackage{subfigure}
\usepackage{wrapfig}
\usepackage{lineno}
\usepackage{algorithm}
\usepackage{algpseudocode}
\usepackage{enumitem}
\usepackage{mathtools}
\usepackage{authblk}
\usepackage{tikz}
\usetikzlibrary{shapes.geometric}
\usepackage{xr-hyper}
\usepackage{setspace}

\usepackage{amsmath,amssymb}
  \usepackage[T1]{fontenc}
  \usepackage[utf8]{inputenc}
\usepackage{lmodern}
\usepackage{xcolor}
\usepackage{graphicx}
\usepackage[breaklinks,colorlinks,linkcolor=black,citecolor=black,urlcolor=black]{hyperref}

\usepackage{dsfont}
\usepackage{comment}

\newtheorem{example}{Example}

\newenvironment{continuedexample}[1]
  {\setcounter{example}{#1}\addtocounter{example}{-1}%
   \renewcommand\theexample{#1 (continued)}%
   \begin{example}}
  {\end{example}}

\newtheorem{assumption}{Assumption}

\newcommand{\E}{\mathbb{E}}
\newcommand{\V}{\mathrm{Var}}
\newcommand{\Cor}{\mathrm{Cor}}

\newcommand{\N}{\mathcal{N}}

\newcommand{\T}{\mathcal{T}}

\newcommand{\g}{\mathrm{g}}

\newcommand{\1}{\mathds{1}}

\DeclareCaptionFont{footnotesize}{\footnotesize}

\definecolor{cbblue}{HTML}{0173b2}
\definecolor{cborange}{HTML}{de8f05}
\definecolor{cbgreen}{HTML}{029e73}
\definecolor{cbred}{HTML}{d55e00}
\definecolor{cbpurple}{HTML}{cc78bc}
\definecolor{cbyellow}{HTML}{ca9161}  

\definecolor{cbgreen2}{HTML}{018571}     
\definecolor{cbblack}{HTML}{252525}   
\definecolor{cbdarkblue}{HTML}{091e75}   
\definecolor{cbbrown}{HTML}{662506}      
\definecolor{cbmagenta}{HTML}{850157}
\definecolor{cbgrey}{HTML}{5c5c5c} 
\definecolor{cbblue2}{HTML}{023eff}  
\definecolor{cborange2}{HTML}{FA8072} 
\definecolor{cbblue3}{HTML}{87CEEB}

\makeatletter
\def\maxwidth{\ifdim\Gin@nat@width>\linewidth\linewidth\else\Gin@nat@width\fi}
\def\maxheight{\ifdim\Gin@nat@height>\textheight\textheight\else\Gin@nat@height\fi}
\makeatother
\setkeys{Gin}{width=\maxwidth,height=\maxheight,keepaspectratio}
\makeatletter
\def\fps@figure{htbp}

\usepackage{geometry}
\usepackage[scaled=0.95]{helvet}  
\usepackage{mathpazo}             

\usepackage{microtype}            
\SetTracking{encoding=*, shape=*}{25}  
\selectfont      
\usepackage[compact]{titlesec}
\titleformat{\section}
  {\normalfont\large\bfseries}{\thesection}{1em}{}
\titlespacing{\section}{0pt}{1.2ex}{0.8ex}
\titlespacing*{\subsection}{0pt}{10pt}{0pt}
\titlespacing*{\subsubsection}{0pt}{10pt}{0pt}

\newcommand{\anon}{1} 

\begin{document}

\def\spacingset#1{\renewcommand{\baselinestretch}%
{#1}\small\normalsize} \spacingset{1}


\if1\anon
{
  \singlespacing
		\title{\bf \hspace{.2cm}
        Sensitivity Analysis for Treatment Effects in Difference-in-Differences Models using Riesz Representation}
			}
        \author[a]{Philipp Bach\footnote{Corresponding author: philipp.bach@fu-berlin.de. This version: \today}}
		\author[b,c]{Sven Klaassen}
		\author[d]{Jannis Kueck}
        \author[d]{Mara Mattes}
        \author[b,c]{Martin Spindler}

        \affil[a]{\it \footnotesize 
        School of Business \& Economics,
        Freie Universität Berlin, Boltzmannstr. 20, 14195 Berlin, Germany}
		\affil[b]{\it \footnotesize Chair of Statistics with Application in Business Analytics, 
                                University of Hamburg Business School, Moorweidenstr. 18, 20148 Hamburg, Hamburg, Germany}
		\affil[c]{\it \footnotesize Economic AI, Nürnberger Str. 262 A, 93059 Regensburg, Bayern, Germany}
		\affil[d]{\it \footnotesize Düsseldorf Institute for Competition Economics, Heinrich Heine University Düsseldorf, Universitätstr. 1, 20225 Düsseldorf, Nordrhein-Westphalen, Germany
		}		\date{}
		
		\maketitle

\begin{abstract}
Difference-in-differences (DiD) is one of the most popular approaches for empirical research in economics, political science, and beyond. Identification in these models is based on the conditional parallel trends assumption: In the absence of treatment, the average outcome of the treated and untreated group are assumed to evolve in parallel over time, conditional on pre-treatment covariates. We introduce a novel approach to sensitivity analysis for DiD models that assesses the robustness of DiD estimates to violations of this assumption due to unobservable confounders, allowing researchers to transparently assess and communicate the credibility of their causal estimation results.
Our method focuses on estimation by Double Machine Learning and extends previous work on sensitivity analysis based on Riesz Representation in cross-sectional settings. We establish asymptotic bounds for point estimates and confidence intervals in the canonical $2\times2$ setting and group-time causal parameters in settings with staggered treatment adoption. Our approach makes it possible to relate the formulation of parallel trends violation to empirical evidence from (1) pre-testing, (2) covariate benchmarking and (3) standard reporting statistics and visualizations. We provide extensive simulation experiments demonstrating the validity of our sensitivity approach and diagnostics and apply our approach to two empirical applications.

\end{abstract}



\section{Introduction} \label{sec:introduction}

Identification of causal effects in difference-in-differences (DiD) models fundamentally relies on the parallel trends assumption. For example, in the canonical $2\times 2$ design with two periods and two groups, it is assumed that, in the absence of treatment, the expected potential outcomes of both groups would have followed parallel trends over time. In practice, however, this assumption is often only plausible after conditioning on observed pre-treatment confounders.   Empirical researchers typically select these covariates based on domain knowledge or economic reasoning relevant to the context of the study. 
The treatment assignment in empirical DiD studies often arises from the decisions of individual units or groups, such as states or countries choosing to adopt certain policies, in response to economic considerations and other factors, some of which may be unobserved.
Empirical researchers try to model these decision processes by accounting for pre-treatment covariates in order to justify parallel trends conditionally on these characteristics. However, identification of the Average Treatment Effect on the Treated (ATT) fails if these variables do not adequately account for all relevant confounding information: The reported ATT estimate will be contaminated by the bias from unobserved self-selection into the treatment groups. 

In such settings, it seems natural to question the validity of the conditional parallel trend assumption: Is the researcher really able to account for all systematic selection mechanisms that make certain types of individuals more likely to receive (or rather choose) the treatment
than others?  Quantifying the implications of such violations can help to assess the robustness of causal findings: If the parallel trend assumption were violated, what would be the resulting bias for the causal estimate? Would this bias be sufficient to substantially change the conclusions of the causal analysis, for example changing the significance of an effect estimate? Such sensitivity considerations are useful for an appropriate interpretation and transparent communication of causal results according to the plausible strength of the parallel trend violation. Building on previous work by \cite{chernozhukov2022long} and \cite{cinelli2020making}, we develop a new approach for sensitivity analysis in DiD models with panel data exploiting the Riesz representation for the ATT in the canonical $2\times 2$ DiD setting and group-time specific average treatment effects, $ATT(\g,t)$ in multi-period settings with staggered adoption. Our sensitivity approach helps to quantify the implications from omitting one or several pre-treatment confounders
in terms of the corresponding explanatory power for the treatment assignment and the observed difference in outcomes over time. Consequently, our framework makes it possible to bound the bias from omitting pre-treatment confounders, adjust the ATT estimators accordingly and to compute critical values, which are also known as ``robustness values'' \citep{cinelli2020making, chernozhukov2022long} or ``breakdown'' values \citep{rambachan2023}.

Our approach builds on the doubly robust ATT estimator introduced by \cite{zimmert2018efficient}, \cite{chang2020double} and \cite{sant2020doubly}, which is compatible with Machine Learning (ML) nuisance estimators in the Double/Debiased Machine Learning (DML) framework \citep{Chernozhukov2018}. 
We derive the Riesz representation for the ATT in the canonical $2\times2$ DiD setting to establish the bias bounds from parallel trends violation and, thus, extend prior work on cross-sectional data by \cite{chernozhukov2022long} and \cite{cinelli2020making}. To relate our sensitivity approach to the common practice of pre-testing in event studies, we also present results for group-time specific average treatment effects in multi-period settings with staggered treatment adoption \citep{callaway2021difference}.

We consider a canonical $2\times 2$ DiD setting with two treatment groups and two periods, in which the conditional parallel trend assumption would be satisfied only if we had access to observed pre-treatment confounders $X_i$ and one or several unobserved confounding variables $A_i$. In general, not taking into account the confounding through $A_i$ will lead to a systematic bias in the estimation of the ATT, which is our target causal parameter $\theta_0$. 
This corresponds to a situation in which researchers worry about the comparability of the treated and control group prior to treatment onset due to differences in pre-treatment characteristics. These would translate into non-parallel trends of the potential outcomes without treatment over the considered evaluation period. A prominent example where individual characteristics play an important role to establish parallel trends is the famous Lalonde data as re-analyzed in \cite{smith2005}. In this example, the individual characteristics of participants in the National Supported
Work (NSW) experiment are not only important to predict whether individuals enter the sample of the treated or control group, but also for the change of the outcome variable (earnings). \cite{smith2005} state that accounting for these time-invariant confounders in a DiD model is crucial to reduce the bias from selection into the treatment groups. In our first empirical example in Section \ref{application}, we apply the suggested sensitivity approach to obtain bias bounds on the ATT in a reevaluation of the data used by \cite{smith2005} and \cite{lalonde1986}. The second empirical application is a replication of \cite{draca2011minimum}, demonstrating the use of pre-testing information for scenarios of parallel trend violations.

Our theoretical results show that the asymptotic bias resulting from violations of parallel trends is closely related to the explanatory power of the unobserved confounding variables $A_i$ for the treatment assignment probability and the outcome difference over time in addition to $X_i$.
We provide asymptotic bounds on the causal parameter $\theta_0$ and corresponding $(1-a)$-confidence limits if we have access to the observed data only, i.e.,  $Y_{i,t},D_{i,t},X_{i}$. For example, this reflects a setting where measurements for pre-treatment confounding are imperfect proxies, such as variables for educational achievement and income capturing confounding from socio-economic status.

A critical step in sensitivity analysis is to define plausible and realistic scenarios of identification violations. A scenario specifies explicit numeric values for the sensitivity parameters, which enter the asymptotic bias formula that is obtained using the Riesz representation for the causal parameters. In our approach, the sensitivity parameters quantify the strength of the parallel trend violation in terms of the explanatory power of $A_i$ for the treatment assignment and the difference in outcomes. We propose three ways to formulate parallel trends violation scenarios: (1) Include information from pre-testing, which is a common practice in event studies with access to pre-treatment periods, (2) exploiting knowledge from leaving-out known and observed pre-treatment confounders (so-called benchmarking) and (3) standard reporting measures and visualizations such as contour plots. In addition to two empirical applications, we also provide evidence on the validity of our sensitivity framework in systematic simulation studies. We would like to highlight that, to the best of our knowledge, the simulation results are the first thorough numerical experiments underscoring the validity of sensitivity analysis based on Riesz representation \citep{chernozhukov2022long}. Our results demonstrate that bias bounds for the confidence intervals achieve near-to-nominal empirical coverage. Moreover, in our simulation experiments, we compare the performance of the sensitivity bounds for the ATT to that of the (infeasible) oracle  estimator for the ATT, which would be obtained if one had access to all observed and unobserved confounders. Our results show that having knowledge on the explanatory power of the unobserved pre-treatment confounder $A_i$ with regard to the treatment status and outcome difference is equivalent to directly having access to the unobserved confounders, already with moderate sample size. We share an open source implementation of our DiD sensitivity analysis with additional documentation and examples through the \texttt{DoubleML} package for Python \citep{DoubleML2022Python, DoubleML}.\footnote{More information on \texttt{DoubleML} available at \url{https://docs.doubleml.org}.}

The remainder of this paper is structured as follows. In Section \ref{literature}, we briefly review the existing literature on difference-in-differences with a focus on violations of the parallel trend assumption, doubly robust estimation, and sensitivity analysis. Section \ref{setting} introduces the difference-in-differences setting and the major idea of our sensitivity approach for DiD. We do so by considering a motivation example, reviewing the underlying ideas of \cite{chernozhukov2022long} and then presenting our new approach in the canonical $2\times 2$ DiD model. Section \ref{sensitivitymulti} extends this framework for sensitivity analysis to the multi-period DiD model with staggered adoption.
Section \ref{simulation} provides simulation studies and Section \ref{application} provides two empirical applications of our proposed framework. Finally, Section \ref{conclusion} concludes and provides an outlook on future research.

\section{Related Literature} \label{literature}

Difference-in-differences is probably the most frequently used approach to causal inference with observational data, see for example recent textbooks by  \cite{chaisemartin2023credible}, \cite{huber2023causal} and \cite{cunningham2021causal}. The recent econometric literature has vastly been impacted by innovative developments in the difference-in-differences literature: New approaches have addressed limitations of classical estimation procedures such as two-way fixed effects (TWFE) in settings with multiple treatment groups and heterogeneous effects, for example including work by \cite{goodman2021difference}, \cite{borusyak2024revisiting}, \cite{sun2021estimating}, \cite{de2020two}, \cite{athey2022design} and \cite{callaway2021difference}. Recent surveys are available in \cite{roth2023}, \cite{chaisemartin2023}, and \cite{callaway2023difference}. A practice-oriented guide to difference-in-differences is provided by \cite{baker2025difference}.
Our study builds on the doubly robust estimator for the ATT in the canonical $2\times 2$ setting as suggested by \cite{sant2020doubly} and on group-time specific average treatment effects, $ATT(\g,t)$, in multi-period designs with staggered adoption as considered by \cite{callaway2021difference}. The latter has been a seminal study to overcome typical limitations of traditional TWFE estimators by recognizing that causal parameters in complex DiD designs can be modeled as aggregations of possibly many $2$-by-$2$ comparisons. Moreover, new estimation approaches have been suggested building on the property of double robustness or Neyman orthogonality. Orthogonality makes it possible to utilize machine learning (ML) for estimation in the double machine learning framework \citep{Chernozhukov2018, sant2020doubly, chang2020double, zimmert2018efficient}.

For a long time, empirical economists and econometricians have recognized the importance of the parallel trend assumption for the causal interpretation of empirical DiD results, which is assumed to hold conditionally or unconditionally on observable characteristics. In the canonical setting with one pre- and one post-treatment period, this assumption states that, in expectation, the potential outcomes under no treatment for the treatment and control group develop in parallel over time. 
However, the parallel trend assumption is untestable. A common practice that is suitable if observations from pre-treatment periods are available, is so-called pre-testing. For example, pre-testing is recommended as part of Pedro Sant’Anna's DiD checklist,\footnote{Available at \url{https://psantanna.com/DiD/checklist.png}.} \cite{roth2023} and in the conclusion of \cite{baker2025difference}.
It is worth noting that the parallel trend assumption makes a statement on the development of the expected potential outcome of the treated group in the post-treatment period if the treatment had not occurred, which is a fundamentally unobservable or counterfactual quantity. The idea of pre-testing is to collect evidence on observable  counterparts of this unknown average potential outcome from pre-treatment periods: If researchers have access to observations prior to the treatment, it is possible to assess whether the observed average difference in the outcome of the treated and control group are the same in these periods. As there is no treatment effective prior to the actual assignment (by ruling out anticipatory effects), the observed outcome difference should be zero. Otherwise, a significant effect would indicate a violation of the parallel trends assumption in the pre-treatment period under consideration. 
Pre-testing serves as a plausibility check rather than a proper statistical test of the (untestable) parallel trends assumption. 
There is no guarantee that evidence suggesting parallel trends prior to the treatment  actually correspond to parallel trends over the considered post-treatment period. The same is true vice versa: Significant pre-trends do not necessarily mean that parallel trends are actually violated after the treatment occurred. In the end, the conclusions from pre-testing exercises have to be interpreted in the specific context of an empirical analysis \citep{baker2025difference}. Recent work by \cite{freyaldenhoven2019pre, bilinski2018nothing, roth2022pretest, kahn2020promise} addresses limitations of pre-testing due to low power. Moreover, \cite{roth2022pretest} point to a risk of selection bias, which arises if researchers only evaluate data for which no pre-treatment violations can be rejected. 

Sensitivity analysis with regard to parallel trend violations is listed as a recommended step in DiD analysis in \cite{baker2025difference}. Despite their relevance, sensitivity approaches to parallel trend violations have only recently been developed. A frequently used approach with a focus on pre-testing has been introduced by \cite{rambachan2023} who extend prior work by \cite{manski2018right}.\footnote{For example, the approach of \cite{rambachan2023} has been used in \cite{callaway2023difference}, \cite{baker2025difference}, \cite{chiu2025}.} \cite{rambachan2023} develop finite-sample and uniformly valid asymptotic bounds when the parallel trends assumption is relaxed. Unlike the point identification of causal parameters under the exact parallel trends assumption, partial identification of a set of causal parameters is obtained under a ``bounded differential trends'' assumption  \citep[p. 153]{chaisemartin2023credible}.  The bounds can be based on user-provided specifications on the shape of the parallel trends violation resulting in a set of different post-treatment trends $\Delta$. The choice of $\Delta$ can be motivated from smoothness assumptions on the differences in trends, pre-testing and their combinations. From a practical perspective, it is appealing that the knowledge gathered from pre-testing can be exploited, for example relative to a multiple of the strongest pre-treatment difference in the parallel-trends in two consecutive periods. By default, this implements a linear extrapolation
of pre-treatment violations over the treatment evaluation periods, which for example can be used to point down so-called breakdown values that correspond to a reduction of the reported effects to zero.
 
An earlier approach to sensitivity analysis with regard to violations of parallel trends is provided by \cite{keele2019patterns}, who adapt previous sensitivity analysis by \cite{rosenbaum2009amplification} and \cite{rosenbaum2002} to DiD designs. In addition, \cite{freyaldenhoven2019pre} develop an approach for identification of the causal parameter in a linear panel setting under violations of the parallel trends assumption. They require an additional identification assumption on the confounding relationships between the unobserved confounders, an additional covariate and the outcome variable: The confounding variables are assumed to affect the additional covariate in a similar way as the outcome, but the treatment variable is not allowed to have an impact on the auxiliary variable. \cite{huber2024joint} develop a joint test for unconfoundedness and conditional parallel trend in a DiD setting based on a testing idea introduced in \cite{huber2023testing}.

In contrast, our approach builds on sensitivity analysis based on Riesz representation as established in \cite{chernozhukov2022long}. A detailed comparison of the framework of \cite{chernozhukov2022long} to Rosenbaum's approach in cross-sectional settings is provided in Appendix E of \cite{chernozhukov2022long}, which similarly applies for the difference-in-differences setting considered here and in \cite{keele2019patterns}. In terms of estimation, the methodology in \cite{keele2019patterns} focuses on a matching approach, whereas we build on the doubly robust estimators of \cite{sant2020doubly} and \cite{callaway2021difference}.

Our approach extends the current literature on difference-in-differences and sensitivity to parallel trends violations in various regards. It provides a new set of tools for analyzing the robustness of DiD estimation results to the existence of unobserved pre-treatment confounders in settings with and without pre-treatment periods. 
\cite{rambachan2023} establish bounds based on a user-provided description of parallel trends violations through a specification of smoothness conditions or relative magnitudes to pre-testing violations, leading to a possibly very flexible set $\Delta$. 
Our bias bounds are motivated by the additional explanatory power of omitted pre-treatment confounding variables relative to the observed variables $X_i$. 
In analogy to the framework of \cite{chernozhukov2022long}, we distinguish two different models: A long and a short model, with corresponding values for the identified (long and short) parameters. Accordingly, the model that contains both observed confounders $X_i$ and unobserved confounders $A_i$ is referred to as the long model. This model would have access to all confounding variables that are sufficient to establish conditional parallel trends and, hence, correspond to what is often called an oracle model. In contrast, the short model only has access to the observable data, i.e., $X_i$, $D_{i,t}$ and $Y_{i,t}$ and thus omits $A_i$. A difference to the sensitivity approach by  \cite{rambachan2023} is that their framework does not assume the existence of such an oracle model. 
Intuitively, our bias bounds are obtained from a systematic comparison of the long and short parameters as identified by the corresponding models using their Riesz representation, whereas \cite{rambachan2023} base their bias bounds on user-provided specification of possible parallel trend violations through $\Delta$, which can be challenging for applied researchers. 
From our point of view, we consider the existence of a long model as plausible and intuitive to applied researchers in many cases, as it often serves as the starting point for identification in DiD models in empirical studies. For example, a common reason for violations to causal identification in economic applications is the existence of socio-economic status (SES).
Usually, empirical researchers employ possibly imperfect proxy measures for SES such as educational attainment, occupation, and income. The sensitivity parameters that we will use to derive the bias from parallel trend violations are defined as (nonparametric) partial $R^2$ values, for example quantifying the share of the residual variation in the outcome difference over time that could be explained by SES in addition to the included and imperfect proxy variables. When applied researchers postulate identification under conditional parallel trends, we believe that this modeling approach often explicitly or implicitly assumes the existence of such an oracle model.

Another difference in our approach compared to \cite{rambachan2023} is the inferential framework, which is built on previous work by \cite{chernozhukov2022long} and \cite{cinelli2020making}. \cite{chernozhukov2022long} provide a general framework for analyzing the omitted variable bias in cross-sectional settings for a wide class of target parameters which can be characterized by a so-called Riesz representation \citep{chernozhukov2022automatic, chernozhukov2021automatic, riesznet}. Accordingly, the estimator of interest is obtained as a solution to a moment equation which can be represented by a linear functional containing two terms: The conditional expectation of the outcome variable and the so-called Riesz representer. 
The latter models the relationship between the covariates and the treatment variable, such as the Horvitz-Thompson transform in augmented inverse probability weighting or Frisch-Waugh-Lovell partialling out of the covariates from the treatment variable in a partially linear regression model.  
We review the general sensitivity framework of \cite{Chernozhukov2018} in Section \ref{setting}. Building on the initial work on linear regression by \cite{cinelli2020making}, \cite{chernozhukov2022long} establish asymptotic bounds on the omitted variable bias and coverage guarantees for confidence bounds in a variety of causal models.

As described before, the bias bounds for the causal parameter and confidence intervals are a function of the sensitivity parameters, which characterize the strengths of the unobserved confounding relationships. An appealing feature based on this modeling approach is that it makes it possible to obtain bounds on the ATT parameter even if no pre-treatment periods are available for pre-testing. In these settings, it would be very difficult to plausibly specify the set of possible parallel trend violations $\Delta$ in the approach by \cite{rambachan2023}. In contrast, our approach allows to leverage the explanatory power of one, several or all observed pre-treatment confounders in our framework to inform violation scenarios. In so-called benchmarking, these variables are left out from the model, which is used to compute empirically grounded values for the sensitivity parameter in the bias formula. Finally, in analogy to the breakdown values in \cite{rambachan2023}, our approach is compatible with the standard reporting statistics of \cite{cinelli2020making} and \cite{chernozhukov2022long}, which inform researchers of how strong unobserved confounding would have to be to cause a reduction of the reported estimate to zero.

Our paper contributes to the existing literature on sensitivity analysis in difference-in-difference models in various regards. First, we establish new results on the asymptotic bias from parallel trends violation for the $ATT$ in the canonical $2\times2$ DiD setting as well as on $ATT(\g,t)$ parameters in multi-period settings with staggered adoption. Our bounds quantify the bias as a function of the explanatory power of the unobserved pre-treatment confounders in terms ot the treatment status and the outcome difference. Second, we propose practical approaches to parameterize the parallel trend violation scenarios. Our approach is compatible with the common practice of pre-testing, but remains applicable even if no-pretreatment periods are available. 
To the best of our knowledge, our third contribution is the first systematic evaluation of sensitivity analysis based on Riesz representation, providing supportive evidence and facilitating the interpretation of its properties, such as the empirical distribution of bias bounds, their comparison to oracle (long) estimates and the empirical coverage of confidence sensitivity bounds. 
Finally, we demonstrate the use of our sensitivity approach in a $2\times2$ DiD setting based on \cite{lalonde1986} and \cite{smith2005} and a multi-period setting in a reassessment of \cite{draca2011minimum}.

\section{Identification, Estimation and Sensitivity Analysis for the ATT 
in the Canonical Difference-in-Differences Design}\label{setting}

In this section, we motivate and introduce our sensitivity approach in the canonical $2\times 2$ DiD  setting with two periods and two treatment groups (treated and control). In Section \ref{sensitivitymulti}, we will extend the sensitivity analysis to group-time average treatment effects in multi-period settings with staggered adoption.

\subsection{Motivating Example}

Before introducing the formal sensitivity framework in the following sections, we briefly illustrate the main ideas in a motivation example.
\begin{figure}
    \centering
    \includegraphics[width=0.8\linewidth]{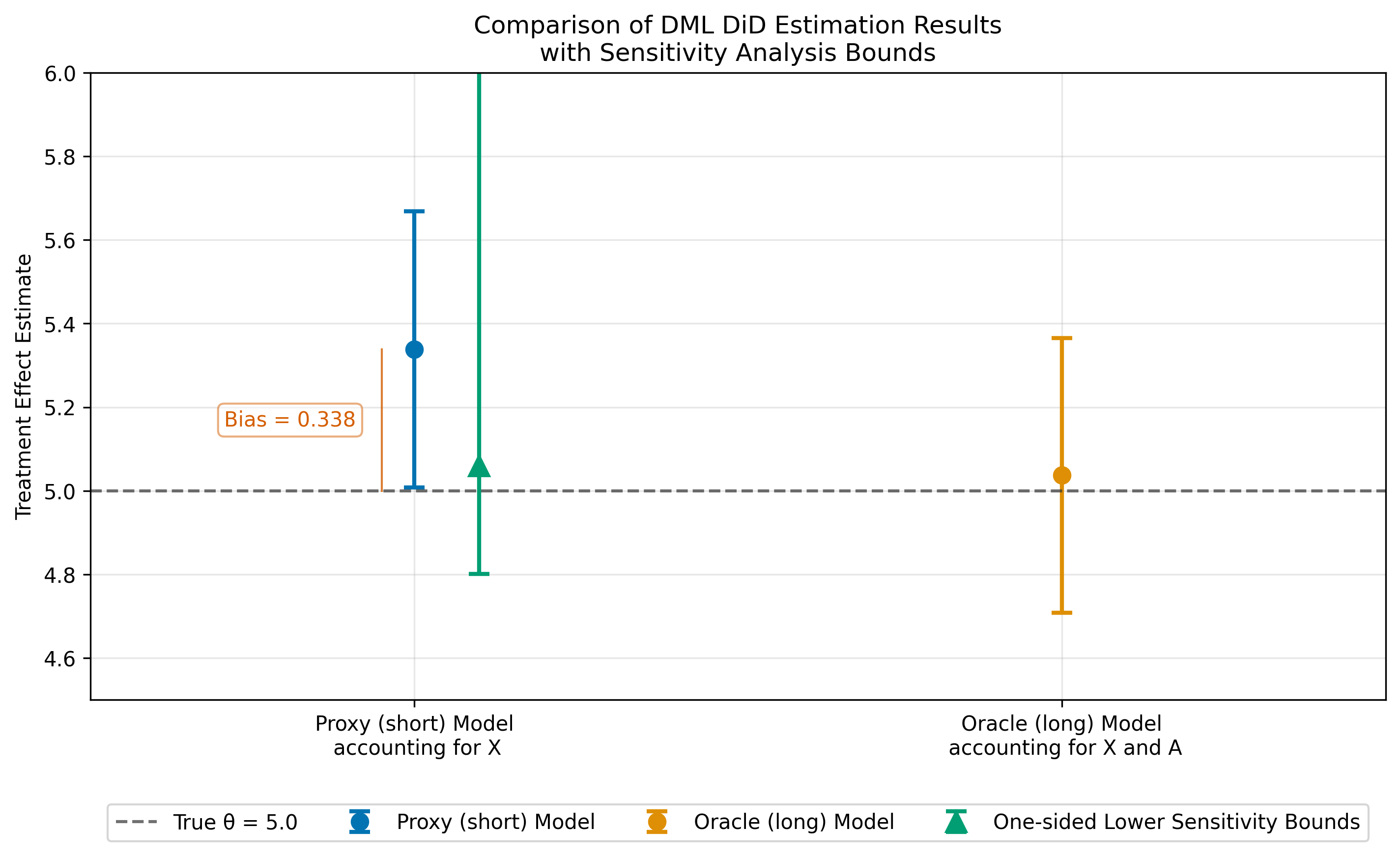}
    \caption{\footnotesize Point estimate and two-sided $90\%$-confidence intervals for the ATT obtained from the short (blue) and long model (orange). The one-sided sensitivity bounds for the point estimator and $90\%$-confidence interval of the short model as obtained from the oracle setting are colored in green. The data is simulated using a modified version of a data generating process adapted from \cite{sant2020doubly}, more information in Section \ref{simulation}. $n=2,500$.}
    \label{fig:motivation}
\end{figure}
We consider the common situation that researchers face in empirical applications of difference-in-differences: Given a set of observed pre-treatment confounders $X_i$, the researcher might worry about unobserved confounding, which is not accounted for by $X_i$. As a consequence, the conditional parallel trend violation would be violated leading to a possibly substantial bias of the causal estimate. 
We can imagine two different models: A feasible model (denoted as the ``short'' model in \cite{chernozhukov2022long})  with access only to the observed pre-treatment confounders $X_i$ and an infeasible (or ``long'') model, which would account for $X_i$ and $A_i$. 
The short and the long model identify the short parameter $\theta_s$ and the long parameter $\theta_0$, which is the true target parameter, respectively. The target parameter in the $2\times 2$ DiD setting is the ATT defined as
\begin{align*}
    \theta_0 = \E\left[Y_{i,1}(1) - Y_{i,0}(0)|D_i=1\right].
\end{align*}
Here, the treatment variable $D_{i,t}=1$ indicates that unit $i$ is treated before time $t$ (otherwise $D_{i,t}=0$). Since $D_{i,0}=0$ for all $i$, we define $D_{i}:=D_{i,1}$. Furthermore, $Y_{i,t}(0)$ denotes the potential outcome of unit $i$ at time $t$ if the unit did not receive treatment up to time $t$ and analogously $Y_{i,t}(1)$ denotes the potential outcome of unit $i$ at time $t$ if the unit did receive treatment. 
Our goal is therefore to make statements on the expected bias of the short parameter as compared to the true ATT,
\begin{align*}
\text{bias}(\theta_0, \theta_s) = \theta_0 - \theta_s.
\end{align*}
As we will show later, the magnitude of this bias will depend on the explanatory power of the unobserved pre-treatment covariates additionally to the observed $X_i$.
Figure \ref{fig:motivation} illustrates the consequences of a parallel trend violation due to omitting the unobserved pre-treatment confounder $A_i$ in the short model. The results show the DML point estimates for the ATT and two-sided $90\%$-confidence intervals obtained from the long (colored orange) and short model (blue) in a simulated data example. The example  is based on a modified version of a data generating process (DGP) from \cite{sant2020doubly}, which is further explained in Section \ref{simulation}. 
In practice, researchers would not be able to know whether the estimated ATT, $\widehat{\theta}_s \approx 5.338$, is close to its true value or not, which is $\theta_0=5.0$ in the example. Applying our suggested approach for sensitivity analysis would result in a lower bound of the ATT of around $\hat{\theta}_{-}=5.059$, which is substantially closer to the true value. Moreover, the standard $90\%$-confidence intervals for the point estimate from the short model do not cover the true ATT. In contrast, we can observe that the one-sided $90\%$-confidence sensitivity bounds derived from our sensitivity approach exhibit coverage of the true effect. The example illustrates that effectively using sensitivity analysis helps to improve the quality of causal statements in DiD studies when researchers are worried about the validity of the parallel trends assumption. As we will later emphasize in our empirical examples, a key ingredient to sensitivity analysis is the proper definition of plausible scenarios of parallel trends violations. In the motivating example, we employed the population values for the underlying sensitivity parameters, which we calibrated in our data generating process. 

\subsection{The Canonical DiD Setting}

We will rely on a notation similar to \citet{sant2020doubly}. Let $Y_{i,t}$ be the outcome of interest for unit $i$ observed at time $t\in\{0,1\}$. 
The observed outcome for unit $i$ at time $t$ is determined by the treatment status,\footnote{Note that the ``switching'' Equation \eqref{eq:switch} for the observed outcome incorporates a no-anticipation assumption stating that $Y_{i,t-1}(1)=Y_{i,t-1}(0)$ which excludes an effect of the treatment variable prior to the realization of $D_i=1$ in period $t$, i.e. $D_{i,t}=1$, see for example \citet{callaway2023difference}.}
\begin{align}
\label{eq:switch}
    Y_{i,t} = D_{i,t}Y_{i,t}(1) +  (1-D_{i,t})Y_{i,t}(0).
\end{align}
Further, let $X_i$ be a vector of observed pre-treatment covariates for unit $i$. Moreover, one or multiple pre-treatment confounders, $A_i$, are unobserved. 
Throughout, we work in a panel setting and assume that the data $W_i = (Y_{i,0}, Y_{i,1}, D_i, X_i, A_i)$ are i.i.d. across units $i \in \{1,\dots,n\}$.
Again, the target parameter of interest is the average treatment effect on the treated (ATT)
\begin{align*}
    \theta_0 = \E\left[Y_{i,1}(1) - Y_{i,0}(0)|D_i=1\right].
\end{align*}
It is useful to define the first difference in observed outcomes over time, $\Delta Y_i := Y_{i,1} - Y_{i,0}$. In the following, we may occasionally abstract from the index $i$ to keep the notation simple. Estimation of the ATT parameter can be based on different approaches, including inverse probability weighting, outcome regression or doubly robust approaches \citep{sant2020doubly, abadie2005semiparametric}.
Our sensitivity approach builds on double machine learning (DML). We define the nuisance parameters $m(x,a)$, denoting the propensity score, and $g(d,x,a)$, denoting the outcome difference conditional on treatment status $d$ and covariates $(x,a)$ as
\begin{align*}
    m(x, a) &:= P(D=1|X=x, A = a)\\
    g(d,x, a) &:= \E[\Delta Y | D=d, X=x, A=a].
\end{align*}

Identification of the ATT is based on the following standard assumptions.
\begin{assumption}[Parallel Trends Conditionally on $X$ and $A$] \label{assumption::parallel_trends}
    $$\E\left[Y_{1}(0) - Y_{0}(0)|D=1, X, A\right] = \E\left[Y_{1}(0) - Y_{0}(0)|D=0, X, A\right] \quad P-a.s.$$
\end{assumption}
\begin{assumption}[Overlap] \label{assumption::overlap}
    There exists an $\epsilon>0$, such that for $p=P(D=1)$, we have
    $$p>\epsilon \text{ and } P(D=1|X,A) = m(x,a) \le 1-\epsilon \quad P-a.s.$$
\end{assumption}
Assumption \ref{assumption::parallel_trends} states that, in expectation, the change in the potential outcomes without treatment would be the same in the treated and untreated group, conditional on all observed and unobserved pre-treatment covariates $X$ and $A$. Identification is therefore compatible with time trends that are specific to the values of $X$ and $A$. Assumption \ref{assumption::overlap} is a commonly imposed overlap assumption that requires a positive fraction of treated individuals and that the propensity score is restricted to values strictly smaller than $1$.

\subsection{Estimation and Inference: Double Machine Learning} \label{sec:dmldid}

We focus on estimation using Double/Debiased Machine Learning (DML) as generally established in \cite{Chernozhukov2018} and adapted to estimation of the ATT in the $2 \times 2$ DiD setting by \cite{chang2020double}, \cite{zimmert2018efficient} and \cite{sant2020doubly}. DML relies on three key ingredients: (1) Neyman orthogonality, (2) high-quality machine learners, and (3) sample splitting. 
Under these three requirements, the DML estimator is consistent and asymptotically normal. Asymptotically valid confidence bounds are provided by \cite{Chernozhukov2018}.
The DML estimator for the ATT is the solution to the orthogonal moment condition and corresponds to the doubly robust estimator in \cite{sant2020doubly}
\begin{align}
\psi(W,\theta,\eta) &= - \frac{D}{p}\theta + \frac{D - m(X,A)}{p(1-m(X,A))}\left(Y_1 - Y_0 -g(0,X,A)\right), \label{eq:dr_score}
\end{align}
with nuisance components $\eta=(m, g)$ being estimated by machine learning. For the sake of notational simplicity, we abstract from in-sample normalization, which is commonly implemented as a finite-sample adjustment. The score function and sensitivity results for the case with in-sample normalization are presented in Appendix \ref{appendix}. Similarly, we abstract from a dedicated notation to highlight out-of-sample predictions, but rather assume that all nuisance predictions are obtained from hold-out partitions from the data to safeguard against overfitting-induced bias \citep{Chernozhukov2018}. 

\subsection{A General Framework for Sensitivity Analysis based on Riesz Representation}  \label{sec:sensitivity-review}

Before we establish new bias bounds for the point estimator of the ATT and confidence intervals in the canonical DiD setting in Section \ref{sensitivity}, we first give a brief review of the general approach for omitted variable bias under violations of the unconfoundedness assumption in cross-sectional settings as established by \cite{chernozhukov2022long}.\footnote{Unconfoundedness is also referred to as \textit{selection-on-observables}, \textit{conditional ignorability} or \textit{conditional exogeneity}.} \cite{chernozhukov2022long} extend previous results on sensitivity analysis for the average treatment effect (ATE) in a linear regression model in \cite{cinelli2020making} to a class of estimators that can be characterized as a linear functional of the conditional expectation of the outcome variable and a so-called Riesz representer,
\begin{align*}
\theta_0 := \E[\mathcal{M}(W,g)] =  \E[g(W) \alpha(W)].
\end{align*}
In the cross-sectional setting, $g(W)$ generally refers to the conditional expectation of the outcome variable. $W$ denotes the data. Furthermore, $\alpha(W)$ is the so-called Riesz representer (RR). The Riesz representer plays a key role for debiasing and implements Neyman orthogonality either through a known analytical expression or through an approximation by an automatic estimation algorithm \citep{chernozhukov2021automatic, chernozhukov2022automatic}. 

\begin{example}[Motivating sensitivity analysis in a partially linear regression model] \label{ex1}

As an illustration, we briefly review the leading example of a partially linear regression model (PLR) from Section 2 in \cite{chernozhukov2022long}.  In this setting, the causal parameter is the coefficient for a continuous treatment variable $D$ in the PLR
model, $Y = \alpha + \theta_0 D + f(X,A) + \nu$ with $\nu$ being an error term and $f(\cdot)$ being a possibly nonlinear function to account for the effect of confounding variables. In this example, we have 
\begin{align*}
\theta_0 = \E[\mathcal{M}(W,g)] &= \E[Y(d+1) - Y(d)] =  \E \left[\E [Y | D = d+1, X,A] - \E[Y|D=d,X,A] \right] \\ &= \E \left[g(d+1,X,A) - g(d,X,A) \right],
\end{align*}
with $Y(d)$ being the potential outcome for the treatment variable being equal to $D=d$ and $g(D,X,A)=\E[Y|D,X,A]$. Hence, $\E[\mathcal{M}(W,g)]$ describes the average treatment effect from increasing $D$ by one unit. 
\end{example}
According to \cite{chernozhukov2022long}, $\theta_0$ can be identified by exploiting the Riesz representation $\E[g(W) \alpha(W)]$. 
\begin{continuedexample}{1}
In the PLR example, the Riesz representation implements the famous Frisch-Waugh-Lovell partialling out of the covariates from the treatment variable,
\begin{align*}
\alpha(W)=\frac{D-\E[D|X, A]}{(D-\E[D|X, A])^2}.
\end{align*}
Hence, under regularity conditions \citep{chernozhukov2022long}, the ATE is identified by
\begin{align*}
\theta_0 = \E \left[(\E [Y | D, X,A]) \left( \frac{D-\E[D|X, A]}{(D-\E[D|X, A])^2} \right) \right].
\end{align*}
\end{continuedexample}
Identification of $\theta_0$ is feasible only if $X$ and $A$ were observed, which, however, is infeasible in an empirical analysis. Hence, \cite{chernozhukov2022long} establish asymptotic bias bounds on the resulting omitted variable bias quantifying the deviation of the short model from the long model. The corresponding quantities for the short model are defined as $g_s(W^s)$ and $\alpha_s(W^s)$ with $W^s=(Y,D,X)$. Accordingly, the bias can be expressed as
\begin{align*}
\text{bias}(\theta_0, \theta_s) = \E[g(W) \alpha(W)]- \E[g_s(W^s) \alpha_s(W^s)].
\end{align*}
Importantly, $\alpha(W^s)$ is the projection of $\alpha$ given the short data,
\begin{align*}
 \alpha_s = \E[\alpha(W) | W^s].
\end{align*}
\cite{chernozhukov2022long} show that the magnitude of the bias depends on the values of the sensitivity parameters, which quantify the difference of the short from the long quantities,
\begin{align}\label{eq:bias}
\text{bias}^2(\theta_0, \theta_s) = \rho^2 C_Y^2 C_D^2 S^2,
\end{align}
with $S^2 = \E[(Y-g_s)^2] \E [\alpha_s^2]$ being a scaling factor that can be estimated empirically and $\rho^2=\text{Cor}^2(g-g_s, \alpha-\alpha_s)$ quantifying the correlation between the residual confounding in the outcome regression and the Riesz representer. For conservative bounds, $\rho$ is set to a value of $\rho=1$. Importantly, the sensitivity parameters $C_Y^2$ and $C_D^2$ quantify the  proportion of residual variance in $Y$ and $\alpha$, respectively, that can be attributed to $A$,
\begin{align*}
C_Y^2&=R^2_{Y-g_s \sim g-g_s}=\frac{\E[(g(W)-g_s(W^s))^2]}{\E[(Y-g_s(W^s))^2]}\\\\
C_D^2 &:=\frac{1 - \frac{\E\big[\alpha_s(W^s)^2\big]}{\E\big[{\alpha}({W})^2\big]}}{\frac{\E\big[\alpha_s(W^s)^2\big]}{\E\big[{\alpha}({W})^2\big]}},
\end{align*}
with $R^2_{V\sim W}=\frac{\V (\E[V|W])}{\V (V)}=\frac{\V(V) - \E[\V(V|W)]}{\V(V)}$ denoting the nonparametric $R^2$ \citep[P. 8]{chernozhukov2022long}. 
The interpretation of the sensitivity parameter $C_Y^2$ is often similar for different causal models and parameters. However, since $C_D^2$ refers to the interpretation of the Riesz representer $\alpha$, which differs across causal models and parameters, the direct interpretation of $C_D^2$ has to be adapted accordingly.
\begin{continuedexample}{1}
In the PLR example, $C_Y^2$ can be interpreted as a nonparametric partial $R^2$ measuring the  proportion of residual variation in the outcome $Y$ that can be explained by $A$ after taking into account the explanatory power of the short model,
\begin{align*}
C_Y^2 = R^2_{Y-g_s \sim g-g_s} &=\frac{\E[\V(Y|X) - \V(Y|X,A)]}{\E[\V(Y|X)]}.
\end{align*}
Similarly, $C_D^2$ in the PLR can be interpreted as the proportion of residual variance in the treatment $D$ explained by $A$,
\begin{align*}
C_D^2=\frac{\mathbb{E}\big[\big(\mathbb{E}[D|X,A] - \mathbb{E}[D|X]\big)^2\big]}{\mathbb{E}\big[\big(D - \mathbb{E}[D|X]\big)^2\big]}.
\end{align*}
\end{continuedexample}
According to Theorem 4 in \cite{chernozhukov2022long}, the DML plug-in estimators for the lower and upper bounds of the target parameter, $\hat{\theta}_{\pm}=\hat{\theta}_s \pm |\rho| C_Y C_D \hat{S}$ are asymptotically normally distributed and centered around their population counterparts, $\theta_{\pm}$. Moreover, the result gives rise to a coverage property of the corresponding asymptotic one-sided confidence bounds, $\ell_-$ and $u_+$, such that $P(\theta_- \ge \ell_-)\rightarrow 1-a$ and $P(\theta_+ \le u_+) \rightarrow 1-a$ given a significance level $a$. Evaluating the bias formula in Equation \eqref{eq:bias} with the oracle values for the sensitivity parameters $\rho$, $C_Y^2$, and $C_D^2$, it is possible to identify the absolute value of the bias. However, these values are generally unknown and researchers have to find plausible, ideally empirically grounded choices to obtain bounds on the bias. These can be obtained from domain expertise and/or benchmarking exercises that take advantage of the explanatory relationships between $Y$, $D$ and the observed covariates $X$ in the data.

\subsection{Sensitivity Analysis for the Canonical Difference-in-Differences Model} \label{sensitivity}

The results in \cite{chernozhukov2022long} refer to identification under the unconfoundedness assumption in cross-sectional setting. In this section, we extend their work to sensitivity analysis with regard to violations of the conditional parallel trend assumption and focus on the case of panel data. Extending our sensitivity approach to repeated cross-sectional data as considered in \cite{sant2020doubly} would be straightforward. This case is covered in the user guide of the \texttt{DoubleML} library for Python with an implementation also provided by \texttt{DoubleML}.\footnote{More information available at \url{https://docs.doubleml.org/stable/guide/sensitivity.html}.} Appendix \ref{appendix} also presents the Riesz representer in the case of in-sample normalization, which is commonly implemented in practice.
We first derive the Riesz representation for the ATT in the canonical $2\times2$ DiD setting. Importantly, other than in the cross-sectional data settings, the conditional expectation in the DiD model with panel data refers to the difference in the outcome over time, $\Delta Y$,
\begin{align*}
    g(d, X, A):= \E[\Delta Y|D=d,X,A],
\end{align*}
where $g(W)=\E[\Delta Y|D,X,A]$ and 
$W=(\Delta Y,X,D,A)$. The Riesz representer for the ATT is given by
\begin{align*}
    \alpha(W) = \frac{D}{p} - \frac{(1-D)}{p}\frac{m(X,A)}{1-m(X,A)},
\end{align*}
with $m(X,A)= E[D|X,A]$ and $p=P(D=1)$.
Hence, the ATT corresponds to
\begin{align*}
    \theta_0 &=
    \mathbb{E}\Big[\mathcal{M}(W,g)\Big]\\
    &= \mathbb{E}\left[\frac{D}{p} (g(1,X,A) - g(0,X,A))\right] \\
    &= \mathbb{E}\left[\frac{D}{p} (g(1,X,A) - g(0,X,A))\right] + \mathbb{E}\left[ g(0,X,A) \left( \frac{D}{p} - \frac{(1-D)}{p} \frac{m(X,A)}{1-m(X,A)} \right)\right] \\
    &= \mathbb{E}\left[ g(D,X,A) \left( \frac{D}{p} - \frac{(1-D)}{p} \frac{m(X,A)}{1-m(X,A)} \right)\right] \\
    &= \mathbb{E}\Big[ g(W) \alpha(W)\Big]
\end{align*}
with $\mathcal{M}(W,g):=D/p (g(1,X,A) - g(0,X,A))$. 
Details are provided in Appendix \ref{appendix}.
It is worth noting that $\theta_0$ is only identified if we would observe $A$ because Assumption \ref{assumption::parallel_trends} is assumed to only hold conditionally on $X$ and $A$.
In practice, we can only work with the observed data $W^s=(\Delta Y,X,D)$. Hence, the short parameter $\theta_s$ is given by 
\begin{align*}
    \theta_s &= 
    \E\Big[\mathcal{M}(W^s,g_s)\Big]
    = \E\left[\frac{D}{p} (g_s(1,X) - g_s(0,X))\right]
=\E\left[ g_s(W^s) \alpha_s(W^s)\right],
\end{align*}
with $g_s(W^s):=\E[\Delta Y|D,X]$. 
Consequently, we can apply the sensitivity methodology from \citet{chernozhukov2022long} and obtain the resulting bounds for the bias
\begin{align*}
|\theta_0-\theta_s|^2\le S^2C_{\Delta Y}^2C_D^2,
\end{align*}
with $S^2:=\E[(\Delta Y-g_s(W^s))^2]\E[\alpha_s^2(W^s)]$.
The interpretation of the corresponding sensitivity parameters
\begin{align*}
    C^2_{\Delta Y} 
    &:= \frac{\E[(g-g_s)^2]}{\E[(\Delta Y-g_s)^2]} = R^2_{\Delta Y -g_s\sim g-g_s}\\
    C^2_D 
    &:= \frac{\E[\alpha^2] - \E[\alpha_s^2]}{\E[\alpha_s^2]} = \frac{1-R^2_{\alpha\sim\alpha_s}}{R^2_{\alpha\sim\alpha_s}},
\end{align*}
leads to interesting results. The strength of confounding generated in the outcome regression $C^2_{\Delta Y}$ directly takes the form
\begin{align*}
    C^2_{\Delta Y}=R^2_{\Delta Y -g_s\sim g-g_s} = \frac{\V\big(\E[\Delta Y|D, X, A]\big) - \V\big(\E[\Delta Y|D, X]\big)}{\V\big(\Delta Y\big) - \V\big(\E[\Delta Y|D, X]\big)},
\end{align*}
which measures the proportion of residual variation in the differenced outcomes $\Delta Y$ that can be explained by the unobserved pre-treatment confounders $A$ in addition to the observed covariates $X$. This directly refers to the effect of violating the conditional parallel trend assumption. 
Furthermore, the strength of confounding generated in the treatment $C^2_D$ can be rewritten as 
\begin{align*}
    C^2_D = \frac{\E\left[\frac{m(X, A)}{1-m(X,A)}\right] - \E\left[\frac{m(X)}{1-m(X)}\right]}{\E\left[\frac{m(X)}{1-m(X)}\right]}.
\end{align*} 
This shows that $C_D^2$ can be interpreted as the relative increase in the average odds of entering the treatment group due to $A$ after taking into account $X$. In the implementation and application of our sensitivity framework, we use a modified version of $C^2_D$ that is bounded by $0$ and $1$ \citep{Bach2024Sensitivity},
\begin{align*}
  \widetilde{C}^2_D:= \frac{C^2_D}{1 + C^2_D} = \frac{\mathbb{E}\left[\frac{m(X, A)}{1-m(X,A)}\right] - \mathbb{E}\left[\frac{m(X)}{1-m(X)}\right]}{\mathbb{E}\left[\frac{m(X, A)}{1-m(X, A)}\right]}.
\end{align*}
Here, $\widetilde{C}^2_D$ can be interpreted as the relative decrease in the average odds of being in the treatment group due to only observing $X$ but not $A$. Again, technical details are provided in Appendix \ref{appendix}. 

The sensitivity parameters emphasize the role of the pre-treatment confounders for identification in difference-in-differences. In order to identify the ATT, it is crucial to account for pre-treatment confounding such that the parallel evolution of the expected potential outcome under no treatment can be justified. The bias from missing information in this regard by not observing $A$ is proportional to the share of the variation in the outcome difference $\Delta Y$ that can be attributed to the unobserved pre-treatment confounding through $A$. For the treatment variable, the confounding variables help to better separate between treated and untreated individuals, as reflected by the odds ratio in the definition of $\widetilde{C}_D^2$. The larger the explanatory power of $A$ for selection into the treatment group, the larger will be the corresponding bias for the ATT.

\section{Sensitivity Analysis in the Multi-Period Difference-in-Differences with Staggered Adoption}
\label{sensitivitymulti} 

In this section, we establish sensitivity analysis for causal parameters in the multi-period DiD setting with staggered adoption as, for example, considered by \cite{callaway2021difference}. 

\subsection{Identification in Difference in Differences with Staggered Adoption}

In the following, we use a notation and identifcation assumptions that are based on \cite{callaway2021difference}. In their paper, \cite{callaway2021difference} show that group-time specific causal effect parameters in a multi-period setting with staggered adoption can be expressed as binary comparisons of the considered treatment and control group. In our presentation, we slightly adjust the notation to better fit into the common naming conventions in the Double/Debiased Machine Learning literature, sometimes slightly abusing notation. The framework is a generalization of the previously presented $2\times2$ DiD setting.  As before, we focus on panel data to abstract from complex notation in the case of repeated cross-sectional data. 

We consider $n$ observational units at time $t = 1, \ldots, \T$.  The treatment status for unit $i$ in period $t$ is indicated by the binary variable $D_{i,t}$. In settings with staggered treatment adoption, it is common to define treatment groups according to the first period after treatment receipt, which requires additional notation. We focus on the case of an absorbing treatment status, i.e., if individual $i$ is treated first in period $\g$, the individual will remain treated until the final period $\T$. The variable $G_i^\g$ is an indicator variable that takes value one if $i$ is treated for the first time in period $\g$, $G_i^\g=\1\{G_i=\g\}$ with $G_i$ referring to the first post-treatment period. In the setting with absorbing treatment, we have $D_{i,t} = 1,$ $\forall t \ge G_i$ almost surely (cf. Assumption 1 in \cite{callaway2021difference}). If individuals are never exposed to the treatment, we define $G_i = \infty$ and $G_i^\g=0, \forall t = 1, \ldots, \T$. We define $Y_{i,t}(0)$ as the potential outcome of individual $i$ in period $t$ if no treatment has been assigned until period $\T$. We summarize the assumptions as follows.

\begin{assumption}[Panel Data]
We assume that the data 
$(Y_{i,1},\dots, Y_{i,\mathcal{T}}, X_i, A_i, D_{i,1}, \dots, D_{i,\mathcal{T}})_{i=1}^n$ are independent and identically distributed (iid).
\end{assumption}

\begin{assumption}[Irreversibility of Treatment]\label{ass:treat_ass}
It holds $D_{i1}=0$ $a.s.$ and $D_{i,t-1}=1$ implies $D_{i,t}=1$ $a.s.$ for all $t=2, \ldots, \T$.
\end{assumption}

The target causal parameters are defined in terms of differences in potential outcomes. The potential outcome of an individual $i$ that has been treated first in period $\g$ can be evaluated in period $t$ by
\begin{align}
Y_{i,t} = Y_{i,t}(0) + \sum_{\g=2}^{\T} \left( Y_{i,t}(\g) - Y_{i,t}(0) \right) G_i^\g.
\label{eq:potentialoutcom
e}
\end{align}

As a measure of the average causal effect of the treatment, it is common to define a group-time average treatment effect parameter, $ATT(\g,t)$. This target parameter quantifies the average change of the potential outcomes due to being treated first in period $\g$ as evaluated in period $t$,
\begin{align} \label{eq:attgt}
ATT(\g,t):= \E \left[Y_{i,t}(\g) - Y_{i,t}(0)  \vert G_i^\g = 1\right].
\end{align}
In the $2\times 2$ DiD setting, the counterfactual average outcome for the treated group (which would have realized had the group not received the treatment) is estimated based on the information from the untreated group. This is valid under the conditional parallel trend assumption, which ensures that conditional on the pre-treatment covariates, the expected average outcome difference over time is the same for the treated and untreated. 
However, in multi-period DiD settings with staggered adoption, there is no longer one unique definition of the control group, whose information can be exploited for identification of the causal parameters. To characterize the control groups in line with the literature, we define an indicator variable $C_{i,t}$, which depends on whether never treated or not yet treated units are used for comparison.
\begin{align*}
C_{i,t}^{(\text{nev})} \equiv C_{i}^{(\text{nev})} &:= 1\{G_i=\infty\} \quad  \text{(never treated)}, \\
C_{i,t}^{(\text{nyt})} &:= 1\{G_i > t\} \quad \text{(not yet treated)}.
\end{align*}
As a consequence, the parallel trend assumption will be adapted to the control group under consideration. To account for anticipation effects, we will first introduce the limited anticipation assumption of \cite{callaway2021difference} using an anticipation parameter $\delta$.

\begin{assumption}[Limited Treatment Anticipation] \label{ass:anticipation}
There is a known $\delta>0$ such that $\mathbb{E}[Y_{i,t}(\mathrm{g})|X_i,A_i, G_i^{\mathrm{g}}=1] = \mathbb{E}[Y_{i,t}(0)|X_i, A_i, G_i^{\mathrm{g}}=1]\ a.s.$ for all $\mathrm{g}\in\mathcal{G}$, $t\in \{1,\ldots, \T\}$ such that $t<\g-\delta$.
\end{assumption}

\begin{assumption}[Parallel Trends Conditionally on $X$ and $A$] \label{ass:cpt}
Let $\delta$ be defined in Assumption \ref{ass:anticipation}. For each $\g \in \mathcal{G}$ and $t\in \{2, \ldots, \T\}$ such that $t\ge \g - \delta$:
\begin{enumerate}
    \item[a.] Never treated control group: 
    $$\mathbb{E}[Y_{i,t}(0) - Y_{i,t-1}(0)|X_i, A_i, G_i^{\mathrm{g}}=1] = \mathbb{E}[Y_{i,t}(0) - Y_{i,t-1}(0)|X_i, A_i,C_{i}^{(\text{nev})}=1] \quad a.s.$$
    \item[b.] Not yet treated control group
    $$\mathbb{E}[Y_{i,t}(0) - Y_{i,t-1}(0)|X_i,A_i, G_i^{\mathrm{g}}=1] = \mathbb{E}[Y_{i,t}(0) - Y_{i,t-1}(0)|X_i,A_i,C_{i,t+\delta}^{(\text{nyt})}=1] \quad a.s.$$
\end{enumerate}
\end{assumption}
Assumption \ref{ass:cpt} a. assumes that, in the absence of treatment, the expected outcome for the group treated first in period $\g$ would have evolved in parallel over the time periods considered as compared to the group that never received the treatment. This assumption is qualitatively different from Assumption \ref{ass:cpt} b., which imposes parallel trends of the group treated first in period $\g$ as compared to groups that have not yet been exposed to the treatment at time $t+\delta$ \citep{callaway2021difference}. Whether never treated or not yet treated are used as a control group depends on the empirical context of an analysis and the overall (causal) research question. Often, only one of these groups is available or relevant for causal evaluations. In addition to the previous assumptions, an overlap assumption has to be made in the multi-period DiD setting to achieve identification of the group-time specific causal parameters.

\begin{assumption}[Overlap] \label{ass:overlap}
For each time period $t=2,\ldots, \T$ and $\g \in \mathcal{G}$, there exists a $\epsilon>0$ such that $P(G^{\g}_i=1)>0$ and $P(G_i^{\mathrm{g}}=1|X_i, A_i, G_i^{\mathrm{g}} + C_{i,t}^{(\text{nyt})}=1) < 1-\epsilon\quad a.s.$
\end{assumption}
Identification of the long parameters, i.e., the $ATT(\g,t)$ in Equation \eqref{eq:attgt}, is granted by  Assumptions \ref{ass:treat_ass} to \ref{ass:overlap} \citep{callaway2021difference}. Again, we note that identification of these target parameters require accounting for $X$ and $A$ in order to justify that the trends in the average outcomes between the defined control and treatment groups develop in parallel over the evaluation period under consideration. 

\subsection{Machine-Learning based Estimation of Group-time Average Causal Parameters} \label{sec:estimation}

It is possible to extend the machine-learning based estimation of the $ATT$ parameter presented in Section \ref{sec:dmldid} to the muli-period DiD model with staggered adoption as presented by \cite{callaway2021difference}. Note that the corresponding nuisance functions depend on the control group used for the estimation of the target parameter. By slight abuse of notation we use the same notation for both control groups $C_{i,t}^{(nev)}$ and $C_{i,t}^{(nyt)}$. More specifically, the control group only depends on
$\delta$ for not yet treated units. 
\begin{align}
g_{0, \mathrm{g}, t_\text{pre}, t_\text{eval}, \delta}(X_i, A_i) &:= \mathbb{E}[Y_{i,t_\text{eval}} - Y_{i,t_\text{pre}}|X_i, A_i, C_{i,t_\text{eval} + \delta}^{(\cdot)} = 1], \label{eq:outcomediff}\\
m_{0, \mathrm{g}, t_\text{eval} + \delta}(X_i, A_i) &:= P \big( G_i^{\mathrm{g}}=1|X_i, A_i, \max(G^{\g}, C_{i,t_\text{eval}}) =1 \big) ,
\end{align}
with nuisance elements $\eta_0=\big(g_{0, \mathrm{g}, t_\text{pre}, t_\text{eval}}, m_{0, \mathrm{g}, t_\text{eval}}\big)$. Here, $g_{0, \mathrm{g}, t_\text{pre}, t_\text{eval},\delta}(\cdot)$ denotes the population outcome change regression function for the control group that is specified to evaluate the causal effect for the group treated first in $\g$  over the pre-period $t_\text{pre}$ and evaluation period $t_\text{eval}$. $t_\text{pre}$ is also denoted as the base period and often specified as the last period before a group received the treatment for the first time, $t_{\text{pre}}= g -\delta - 1$. 
Furthermore, $m_{0, \mathrm{g}, t_\text{eval} + \delta}(X_i, A_i)$ is the generalized propensity score. For notational purposes, we will omit the subscripts $\g, t_{\text{pre}}, t_{\text{eval}},\delta$ and refer to the corresponding functions by the simplified versions 
\begin{align*}
g_0(0, X_i, A_i) &\equiv g_{0, \g, t_{\text{pre}}, t_{\text{eval}}, \delta}(X_i, A_i) \\
m_0(X_i,A_i)&\equiv m_{0, \g, t_{\text{eval}} + \delta}(X_i,A_i).
\end{align*} 
Note that for estimation of the causal parameters in DiD models, it suffices to estimate the conditional expectation of the outcome differences for the specified control group as indicated by conditioning on $C_{i,t_{\text{eval}+\delta}}=1$ in Equation \eqref{eq:outcomediff}. However, for sensitivity analysis, we require additional estimation of the conditional outcome difference for the group treated first in period $\g$, which we define as
\begin{align*}
g_0(1, X_i, A_i)\equiv \mathbb{E}[Y_{i,t_\text{eval}} - Y_{i,t_\text{pre}}|X_i, A_i, G_i^{\mathrm{g}} = 1]
\end{align*}
machine-learning based estimation of the $ATT(\g,t)$ parameters can be based on the multi-period analog of the Neyman-orthogonal score for the $ATT$ in the canonical DiD setting in Equation \eqref{eq:dr_score},
\footnote{Again, we abstract from in-sample normalization in our presentation. The score function and Riesz representation with in-sample normalization are presented in Appendix \ref{appendix}.}
\begin{align}
\begin{aligned} \label{eq:dr_score_multi}
\psi(W,\theta,\eta) &= - \frac{G^{\mathrm{g}}}{\mathbb{E}[G^{\mathrm{g}}]}\theta + \frac{G^{\mathrm{g}} - m(X,A)}{\mathbb{E}[G^{\mathrm{g}}](1-m(X,A))}\left(Y_{t_\text{eval}} - Y_{t_\text{pre}} - g(0,X,A)\right).
\end{aligned}
\end{align}

\subsection{Sensitivity Analysis in a Multi-Period DiD Setting}

To extend the sensitivity framework from Section \ref{sensitivity}, we derive the Riesz representer for the multi-period DiD setting with staggered adoption. 
In this setting, the moment equation is
\begin{align}
\begin{aligned}\mathcal{M}(W,g) &= \big(g(1,X,A) - g(0,X,A)\big)\cdot \frac{G^{\g}}{\E[G^{\g}]}\cdot \max(G^{\g}, C^{(\cdot)}),  \label{eq:moment-multi}
\end{aligned}
\end{align}
where $\max(G^{\g}, C^{(\cdot)})$ is an indicator that takes value one if an individual belongs to the treated or the specified control group. Including this indicator ensures that only observations from the relevant treatment and control groups are used for identification. When we later present the sensitivity parameters, it is important to account only for variation in the data that is relevant for the target causal parameter.  
This gives rise to the Riesz representer for the $ATT(\g,t)$
\begin{align*}
\alpha(W) = \left(\frac{G^{\g}}{\E[G^{\g}]} - \frac{m(X,A)(1-G^{\g})}{\E[G^{\g}](1-m(X,A))}\right) \cdot \max(G^{\g}, C^{(\cdot)}).
\end{align*}
To have a compact presentation of the sensitivity parameters we refer to the outcome difference that is estimated for evaluation of the $ATT(g,t)$ parameter by $\Delta_{t_{\text{pre}},t_{\text{eval}}}Y = Y_{t_{\text{eval}}} - Y_{t_{\text{pre}}}$.

Accordingly, the sensitivity parameters are given by 
\begin{align*}
     & C^2_{\Delta Y}  =  R^2_{\Delta_{t_{\text{pre}},t_{\text{eval}}} Y -g_s\sim g-g_s} =  \\ \\
& \frac{\mathrm{Var}\big(\E[\Delta_{t_{\text{pre}},t_{\text{eval}}} Y|D, X, A, \max(G^{\g}, C^{(\cdot)}) =1 ]\big) - \mathrm{Var}\big(\E[\Delta_{t_{\text{pre}},t_{\text{eval}}} Y |D, X, \max(G^{\g}, C^{(\cdot)}) =1 ]\big)}{\mathrm{Var}\big(\Delta_{t_{\text{pre}},t_{\text{eval}}} Y | \max(G^{\g}, C^{(\cdot)}) =1\big) - \mathrm{Var}\big(\E[\Delta_{t_{\text{pre}},t_{\text{eval}}} Y|D, X, \max(G^{\g}, C^{(\cdot)}) =1]\big)},
\end{align*}
and, for the selection into the treatment groups,
\begin{align*}
C_D^2 &=\frac{\E\left[ \frac{m(X,A)}{1-m(X,A)} \lvert \max(G^{\g}, C^{(\cdot)}) =1 \right] - \E\left[\frac{m(X)}{1-m(X)} \lvert \max(G^{\g}, C^{(\cdot)}) =1 \right] }{\E\left[\frac{m(X)}{1-m(X)} \lvert \max(G^{\g}, C^{(\cdot)}) =1 \right] }.
\end{align*}
As before, it is useful to use the transformed version $\widetilde{C}_D^2$.
\begin{align*}
\widetilde C^2_D := \frac{C^2_D}{1+C^2_D}=\frac{\E\left[ \frac{m(X,A)}{1-m(X,A)}\lvert \max(G^{\g}, C^{(\cdot)}) =1 \right] - \E\left[\frac{m(X)}{1-m(X)} \lvert \max(G^{\g}, C^{(\cdot)}) =1 \right] }{\E\left[\frac{m(X,A)}{1-m(X,A)} \lvert \max(G^{\g}, C^{(\cdot)}) =1\right] }.
\end{align*}
The sensitivity parameters quantify the explanatory power of the unobserved confounders for the considered outcome difference and the treatment assignment, which is analguous to their interpretation in the $2\times 2$ case. However, we have to restrict attention to only those observations in the data that are relevant for the binary comparison in the multi-period DiD setting. This is done by conditioning on the indicator $\max(G^\g, C^{(\cdot)})$. 

$C^2_{\Delta Y}$ is then measuring the share of the residual variation in the difference of the outcome variable over the period $t_{\text{pre}}$ to $t_{\text{eval}}$ for the group that received treatment first in period $\g$, $\Delta_{t_{\text{pre}},t_{\text{eval}}}Y$, which can be explained by accounting for the unobserved pre-treatment confounders $A$. $\widetilde C^2_D$ measures the relative decrease in the odds ratio to receive the treatment for the first time in period $\g$ due to not accounting for the explanatory power of $A$.

\subsection{Determining Scenarios with Parallel Trend Violations} \label{sec:scenarios}

The previous sections provided a bias formula for causal parameters in difference-in-differences models  to quantify violations of the conditional parallel trend assumption. Once we plug in values for $C_D^2$ and $C_{\Delta Y}^2$, we obtain an upper and lower bound for the ATT and the $ATT(g,t)$ parameters, respectively. Moreover, we can exploit Theorem 4 of \cite{chernozhukov2022long}, which makes it possible to establish coverage guarantees of the confidence bounds.  Heuristically speaking, plugging in the population values for the sensitivity parameters in the bias formula in Equation \eqref{eq:bias} is asymptotically equivalent to having direct access to the $A$ in the oracle model.

In empirical applications, researchers cannot use the long data $W=(Y,D,X,A)$, nor can they know the population values for the sensitivity parameters. As they are concerned with the presence of unobserved pre-treatment confounders $A$, they are faced with choosing values for $C_D^2$, $C_{\Delta Y}^2$, and $\rho$ that are plausible for their specific empirical data setting and, ideally, close to the population values of the sensitivity parameters.\footnote{It is generally recommended to start with the most conservative value $\rho=1$.} We denote the choice of specific values for $C_D^2$ and $C_{\Delta Y}^2$ as parametrizing a \textit{violation scenario of the conditional parallel trend assumption} (or shorter, simply \textit{violation scenarios}), for which the asymptotic bias is estimated. We suggest four practical ways, in which empirical researchers can obtain plausible values for $C_D^2$ and $C_{\Delta Y}^2$: First, domain expertise can help to quantify the plausible explanatory power of unobserved pre-treatment confounders for $\Delta_{t_{\text{pre}},t_{\text{eval}}}Y$ and $G^\g$. Second, standard reporting and visualizations can serve as informative measures, which might be complementary to domain expertise. Third, information from pre-treatment periods might be exploited, following the rationale of pre-testing in event studies. Fourth, so-called benchmarking, i.e., leaving out observed pre-treatment confounding variables, helps to obtain empirically grounded choices for $C_D^2$ and $C_{\Delta Y}^2$. In the end, the definition of one or more violation scenarios can be the result of combining these four procedures.

\subsubsection*{Standard Reporting and Visualization}

\cite{cinelli2020making} and \cite{chernozhukov2022long} develop a suite of standard reporting measures in their sensitivity framework, which can also be applied in difference-in-differences models. The so-called \textit{robustness values} $RV$ and $RV_a$ indicate critical scenarios of violations of the identifying assumptions: In the difference-in-differences setting, $RV$ corresponds to the minimum strength of violating the conditional parallel trends assumption that suffices to set the ATT or, respectively, the $ATT(\g,t)$ estimate equal to zero or an alternatively chosen null hypothesis. $RV_a$ accounts for estimation uncertainty and reports the minimum violation scenario that is sufficient to make the causal parameter non-significant, hence $RV \ge RV_a$ in general. Note that the robustness values always refer to symmetric violation scenarios, $C_{\Delta Y}^2 = \widetilde C_D^2$. The bias bounds for the point estimate or the confidence limits can be illustrated in contour plots, as for example presented in Figure \ref{fig:contour_draca} in Section \ref{application}. A contour line indicates all combinations of $C_{\Delta Y}^2$ and $\widetilde C_D^2$ that correspond to the same value of the bias-adjusted estimate.

\subsubsection*{Pre-testing}

Pre-testing is a common practical procedure to assess violations of parallel trends in difference-in-differences models with multiple periods. Whereas no formal statistical test exists for testing a violation of parallel trends over the relevant time periods for estimation of the causal parameter, the underlying idea is to provide evidence from pre-treatment periods. In absence of an actual treatment, a significant causal effect for one or several pre-treatment comparisons might cast doubt on the validity of conditional parallel trends in the relevant treatment periods. However, not rejecting the null hypothesis of a zero effect in the pre-treatment period does not serve as evidence for the validity of the identification assumptions. In the end, the conclusions from pre-testing evidence are dependent on the context of the empirical application.

It is possible to use pre-testing information to empirically support the choice of sensitivity parameters $C_D^2$ and $C_{\Delta Y}^2$. Here, we focus on sensitivity analysis for the average treatment effect for one treatment group that has received the treatment first in period $\g$ being evaluated in the first post-treatment period, 
i.e., $ATT(\g, \g)$ with pre-treatment period $\g-1$. Suppose, we have access to $\T_{\text{pre}}$ pre-treatment periods, i.e, we have periods $1,\ldots, \T_{\text{pre}}, \g, \g+1, \ldots, \T$. Pre-testing makes it possible to obtain $\T_{\text{pre}}-1$ placebo effects $\hat{\theta}_{2}, \ldots, \hat{\theta}_{\T_{\text{pre}}}$. Because there is no treatment in the placebo periods and anticipatory effects are ruled out by assumption, any measured pre-testing effect is the result of a parallel trends violation \citep{roth2022pretest, rambachan2023}. Hence, it is possible to bound the bias in the pre-treatment periods by
\begin{align*}
\lvert \theta_{0,t} - \hat{\theta}_{t}\lvert ^2 = \lvert 0 - \hat{\theta}_{t}\lvert ^2 = \lvert \hat{\theta}_t \lvert ^2, \quad \forall t \in \{2, \ldots, \T_{\text{pre}}\}.
\end{align*}
The robustness value for each pre-treatment period $t$, $RV_t$, indicates the values $C_{\Delta Y}^2$ and $\widetilde C_D^2$ that would suffice to reduce (or rather correct) the bias from a conditional parallel trend violation to zero. In our application in Section \ref{application}, we employ a conservative rule that takes the maximum robustness values from all pre-treatment periods (or a $k$-fold multiple thereof, $k>0$):\footnote{It might be that the sign of the pre-treatment estimates is different than that of the suspected bias. In that case, the rule might be adjusted to select the maximum among all pre-treatment violations in the suspected direction of the bias.}
\begin{align*}
C_{\Delta Y}^{2, \text{pre}}=\widetilde C_D^{2, \text{pre}} =\max \{RV_2, \ldots, RV_{\T_{\text{pre}}}\}.
\end{align*}

\subsubsection*{Benchmarking: Leaving out known pre-treatment confounders}

An additional way to obtain specific values for the sensitivity parameters, which is applicable also if no pre-treatment periods are available, is so-called benchmarking. Leaving out observed confounding variables has previously suggested for example in \cite{cinelli2020making} and \cite[Appendix F]{chernozhukov2022long}  who build on prior work by \cite{imbens2003sensitivity}, \cite{altonji2005selection}, and \cite{oster2019unobservable}. Adapting the idea of leaving-out-confounders from the cross-sectional setting, we can mimic the omission of unobserved pre-treatment confounders by leaving out one or multiple variables from $X$. Whereas there is no formal guarantee that the actual confounding variable $A$ shares the same explanatory power with the benchmarking variable $X_j$, these modeling exercises are very useful for empirical researchers: In many cases, domain experts and empirical researchers know about the role of the most important pre-treatment covariates, which are crucial to justify the conditional parallel trend assumption. Leaving these variables out can be helpful to judge the plausibility of violation scenarios based on the empirical evidence.

Benchmarking works as follows: Let $X_j$ denote the benchmarking pre-treatment covariates, leaving only $X_{-j}=X \setminus \{X_j\}$ for estimation of the nuisance functions $g(\cdot)$ and $m(\cdot)$. Then it is possible to recompute the sensitivity parameters that correspond to the change in the estimate of the ATT or, respectively, $ATT(\g,t)$ that is caused by removing $X_j$ from the proxy model. We denote these calibrated sensitivity parameters as $\widetilde C_D^{2,\text{bench}}$ and $C_{\Delta Y}^{2,\text{bench}}$. It is necessary to incorporate a correction factor $\kappa$ that adjusts for the change in the residual variation that is left after removing $X_j$ from $X$ in the proxy model, see also Appendix F of \cite{chernozhukov2022long}.

\section{Simulation Study}\label{simulation}

In this section, we report the results of a simulation study to assess the finite-sample performance of our sensitivity framework. To the best of our knowledge, these are the first systematic simulation results for sensitivity analysis based on Riesz representation. We extend a data generating process (DGP) from \cite{sant2020doubly} for the canonical $2\times 2$ DiD setting in terms of an unknown confounder $A$. As we will clarify in the following, we implement a confounding setting with known population values for the sensitivity parameters $\rho, C_{\Delta Y}^2$, and $\widetilde C_D^2$. This makes it possible to assess the empirical performance of our sensitivity approach in four regards: First, given the oracle values for the sensitivity parameters, we can estimate the lower and upper bias bounds for the ATT as obtained from Equation \eqref{eq:bias}, $\hat{\theta}_-$ and $\hat{\theta}_+$, and compare them to the true parameter value, $\theta_0$. In the scenario considered, the parallel trend violation implements an upward bias of the short ATT parameter, such that we focus on the evaluation of the lower bias bound. Second, we can evaluate the empirical performance of the robustness values $RV$ and $RV_a$, which are expected to be close to the oracle values of $C_{\Delta Y}^2$, and $\widetilde C_D^2$.\footnote{We implemented a symmetric parallel trend violation scenario, such that the population values are calibrated to $C_{\Delta Y}^2=\widetilde C_D^2=0.1$, which implements a nominal level for the robustness values $RV$ and $RV_a$ of $0.1$ .} Third, we can evaluate the performance of the lower and upper bias bounds for the confidence limits, $\hat{\ell}_-$ and $\hat{u}_+$. Evaluating the sensitivity bounds at the population values for $C_{\Delta Y}^2$ and $\widetilde C_D^2$, the one-sided bounds are expected to cover the true ATT in $1-a$ percent of the data realizations. Lastly, we can compare the bias bounds to the long estimate of the ATT, $\hat{\theta}_{\text{long}}$, which is only feasible in a simulation setting. Unlike in real data analysis, we can use the simulated pre-treatment confounder $A$ to compute the DML estimate that would be obtained from the long data. This is informative to assess the usability of the sensitivity bounds according to the sensitivity parameters as compared to having access to the unobserved pre-treatment confounder $A$.

\subsection{Data Generating Process}

We consider an adapted version of the Monte Carlo simulation considered in \citet{sant2020doubly}. Let $X=(X_1, X_2, X_3, X_4, X_5)^T \sim \mathcal{N}(0,\Sigma),$ where $\Sigma$ corresponds to the identity matrix. For $j=1,2,3,4,5$, define $Z_j=(\widetilde{Z}_j-\E_n[\widetilde{Z}_j])/\V(\widetilde{Z}_j)$, where
\begin{align*}
\begin{aligned}
    \widetilde{Z}_1 &= \exp(0.5\cdot X_1)   &\qquad \widetilde{Z}_2 &= 10 + \frac{X_2}{1 + \exp(X_1)} \\
   \widetilde{Z}_3 &= \left(0.6 + \frac{X_1 \cdot X_3}{25} \right)^3 & \qquad  \widetilde{Z}_4 &= (20 + X_2 + X_4)^2 \\
      \widetilde{Z}_5 &= X_5 && \\
\end{aligned}
\end{align*}
For generic $V=(V_1, V_2, V_3, V_4, V_5)^T$, define
\begin{align*}
    f_{\text{reg}}(V) &= 210 + 27.4\cdot V_1 + 13.7\cdot (V_2 + V_3 + V_4)\\
    f_{\text{ps}}(V) &= 0.75\cdot (-V_1 + 0.5\cdot V_2-0.25\cdot V_3 - 0.1\cdot V_4).
\end{align*}
Using only the observed pre-treatment confounders $X$ or, respectively, $Z$ in the population model would basically implement the DGP from \cite{sant2020doubly}. However, we extend the simulation design in terms of an unobserved pre-treatment confounder $A$, which is uniformly distributed over an interval $(-1,1)$, i.e., $A\sim \mathcal{U}(-1,1)$. $A$ enters the equation of the propensity score and the outcome difference regression in an additive way\footnote{Additivity in the propensity score helps to compute the population values for the Riesz representer. To ensure that $p(Z,A)\in (0,1)$, we impose an additional clipping such that $0.1 \le p(Z,A) \le 0.9$.}
\begin{align*}
    p(Z, A) &= \frac{\exp(f_{\text{ps}}(Z))}{1 + \exp(f_{\text{ps}}(Z))} + \gamma_A \cdot A\\
    D &= \1\{p(Z, A) \ge U\},
\end{align*}
with $U \sim\mathcal{U}[0,1]$.
The outcome $Y$ is generated as
\begin{align*}
    Y_0(0) &= f_{\text{reg}}(Z) + \beta_A \cdot A +\varepsilon_0\\
    Y_1(D) &=   D\cdot \theta\cdot (Z_5 + 1) + f_{\text{reg}}(Z) + \beta_A \cdot A + \varepsilon_1(D),
\end{align*}
where $\varepsilon_0,\varepsilon_1(D)\sim\N(0,\sigma_\varepsilon^2)$ and $\theta\in \mathbb{R}$. 

\begin{table}[t]
\centering
\begin{tabular}{lccccc}
\toprule
$n$ & $\hat{\theta}_s$ & $\hat{\theta}_{-}$ & $\hat{\theta}_{+}$ & $\hat{\theta}_{\text{long}}$ & $RV_{\theta=5}$ \\
\midrule
500 & 5.301 (0.424) & 5.001 (0.430) & 5.600 (0.423) & 4.997 (0.413) & 0.134 (0.092) \\
1000 & 5.303 (0.289) & 5.004 (0.290) & 5.602 (0.290) & 4.995 (0.284) & 0.113 (0.072) \\
5000 & 5.306 (0.129) & 5.005 (0.129) & 5.606 (0.129) & 5.001 (0.126) & 0.101 (0.040) \\
10000 & 5.305 (0.091) & 5.005 (0.091) & 5.605 (0.091) & 5.001 (0.090) & 0.101 (0.029) \\
50000 & 5.303 (0.041) & 5.003 (0.041) & 5.604 (0.041) & 4.999 (0.040) & 0.101 (0.013) \\
\bottomrule
\end{tabular}
\caption{ Average simulation results for point estimation based on $10,000$ replications and $\theta_0=5.0$. Estimation and bias bounds for ATT, standard deviations in brackets. $\hat{\theta}_s$: DML estimate for ATT under parallel trend violation (short model); $\hat{\theta}_-$ and $\hat{\theta}_+$: Lower  and upper bound for ATT according to bias adjustment with population sensitivity parameters; $\hat{\theta}_{\text{long}}$: DML estimate as obtained from using long data, including pre-treatment confounder $A$; $RV_{\theta=5}$: Robustness value with null hypothesis $\theta=5$, nominal value in the simulation design is $0.1$.}
\label{tab:res1}
\end{table}
We parametrize the causal model such that for the ATT we have $\theta_0 = 5$. Simulation studies for sensitivity analysis are characterized by a specific challenge. We would like to implement a specific confounding scenario, which is in line with the previously presented theoretic framework. To do so, we calibrate the values for $\gamma_A$ and $\beta_A$ based on a super-population model with $1,000,000$ observations, for which we can compute the long and short model. Accordingly, we can compute the population-level values of the sensitivity parameters $\widetilde C_D^2$, $C_{\Delta Y}^2$, and $\rho$. These values are then used as the evaluated parallel trend violation scenario in the empirical application of the sensitivity framework, such that we can measure the performance of the corresponding bias bounds at these population values. We consider settings with sample size $n \in \{500, 100, 5000, 50000\}$ and report results from $R=10,000$ simulation repetitions. We consider specification for the outcome regression and propensity scores that are linear in the covariates, i.e., $Z=\widetilde{Z}$, which corresponds to DGP 1 in \cite{sant2020doubly}. For estimation, we use unpenalized linear and logistic regression learners. Hence, the resulting bias of the ATT estimate will be only the consequence of the parallel trends violation and not reflect misspecification bias.

\subsection{Results}

Table \ref{tab:res1} shows the average results for the point estimation and bias bounds for the ATT as obtained from $R=10,000$ simulation repetitions for different sample sizes $n$. In all settings, the short estimate of the ATT, $\hat{\theta}_s$, exhibits an upward bias that results from omitting $A$ from the model. Applying the bias formula in Equation \eqref{eq:bias} according to the definitions of the sensitivity parameters presented in the previous sections leads to a lower bound, $\hat{\theta}_-$, that is very close to the true value $\theta_0=5.0$. In all settings, the robustness value is close to its nominal level of $RV_{\theta=5}=0.1$. In the setting with the smallest sample size, $n=500$, the robustness value is slightly too optimistic suggesting to use conservative settings in small-sample settings. This result can be explained by some numerical instabilities in small samples, which is partly reflected by the higher standard variation. We recommend considering estimation uncertainty when interpreting robustness values, as also reflected by low values for the robustness values $RV_{\theta=5,a=0.1}$ in these settings, cf. Table \ref{tab:res2}. Interestingly, comparing the lower bound, $\hat{\theta}_-$, to the oracle estimator $\hat{\theta}_{\text{long}}$ reveals that using the oracle confounding scenario is almost equivalent to directly using the unobserved pre-treatment confounder $A$. In small samples, the bias bounds are slightly more variable than the oracle estimate. However, with increasing sample size, this difference becomes negligible. 

\begin{table}[t]
\begin{tabular}{lccccccc}
\toprule
n & $\hat{\ell}_s$ & $\hat{\ell}_{-}$ & $\hat{\ell}_{\text{long}}$ & $\theta_0 \ge \hat{\ell}_s$ & $\theta_0 \ge \hat{\ell}_{-}$ & $\theta_0 \ge \hat{\ell}_{\text{long}}$ & $RV_{\theta=5,a=0.1}$ \\
\midrule
500 & 4.688 (0.438) & 4.388 (0.444) & 4.395 (0.417) & 0.768 & 0.926 & 0.928 & 0.019 (0.046) \\
1000 & 4.877 (0.290) & 4.578 (0.291) & 4.577 (0.284) & 0.660 & 0.927 & 0.932 & 0.022 (0.041) \\
5000 & 5.117 (0.129) & 4.817 (0.129) & 4.816 (0.126) & 0.178 & 0.923 & 0.929 & 0.044 (0.036) \\
10000 & 5.172 (0.091) & 4.872 (0.091) & 4.870 (0.090) & 0.029 & 0.921 & 0.925 & 0.058 (0.029) \\
50000 & 5.244 (0.041) & 4.943 (0.041) & 4.941 (0.040) & 0.000 & 0.919 & 0.932 & 0.082 (0.013) \\
\bottomrule
\end{tabular}
\caption{ Average simulation results for confidence limits for $10,000$ replications and $\theta_0=5.0$. Estimation and bias bounds for lower confidence limit, standard deviations in brackets. $\hat{\ell}_s$: Lower limit of one-sided $90\%$-confidence interval from DML inference under parallel trend violation (short model); $\hat{\ell}_{-}$: Lower one-sided confidence bias bound at nominal level $90\%$; $\hat{\ell}_{\text{long}}$: Lower limit of one-sided $90\%$-confidence interval from DML inference using long data, including pre-treatment confounder $A$; $RV_{\theta=5, a=0.1}$: Robustness value accounting for estimation uncertainty at significance level $0.1$ and null hypothesis $\theta=5$, nominal level in the design is $0.1$.}
\label{tab:res2}
\end{table}
Table \ref{tab:res2} shows the average results for estimation and sensitivity bounds of the lower confidence limit. The results illustrate that, in line with the expectations, the empirical coverage of the one-sided confidence interval $[\hat{\ell}_s,\infty)$ for the ATT is below the nominal level of $90\%$ and approaches $0.00$ as estimation uncertainty diminishes in larger samples. In contrast, the lower confidence bias bounds, $\hat{\ell}_-$, cover the true value of the ATT in $91.9\%$ to $92.7\%$ of the cases, approaching a nominal coverage of $90\%$. In larger sample settings, the lower bound of the confidence limit gets closer to the true value for the ATT.\footnote{Note that the population lower bound, $\theta_-$, evaluated in the population parallel trend violation scenario is approximating the true ATT, $\theta_0$ (subject to numerical differences).} The robustness value $RV_{\theta=5,a=0.1}$ appears to be conservative in small samples but approaches the nominal value of $10\%$ in larger samples. Comparing the performance of the lower confidence bias bound to that of the lower confidence corresponding to the long point estimator reveals that their performance is similar in terms of empirical coverage and variability. The average value for $\hat{\ell}_-$ is very close to $\hat{\ell}_{\text{long}}$, with differences becoming smaller in larger samples. Moreover, increasing the sample size leads to smaller variability of the bias bounds, such that the corresponding standard deviations approach those of the oracle confidence bound in moderate and large samples.

\begin{figure}[t]
    \centering
    ($i$)\\
    \includegraphics[width=0.6\linewidth]{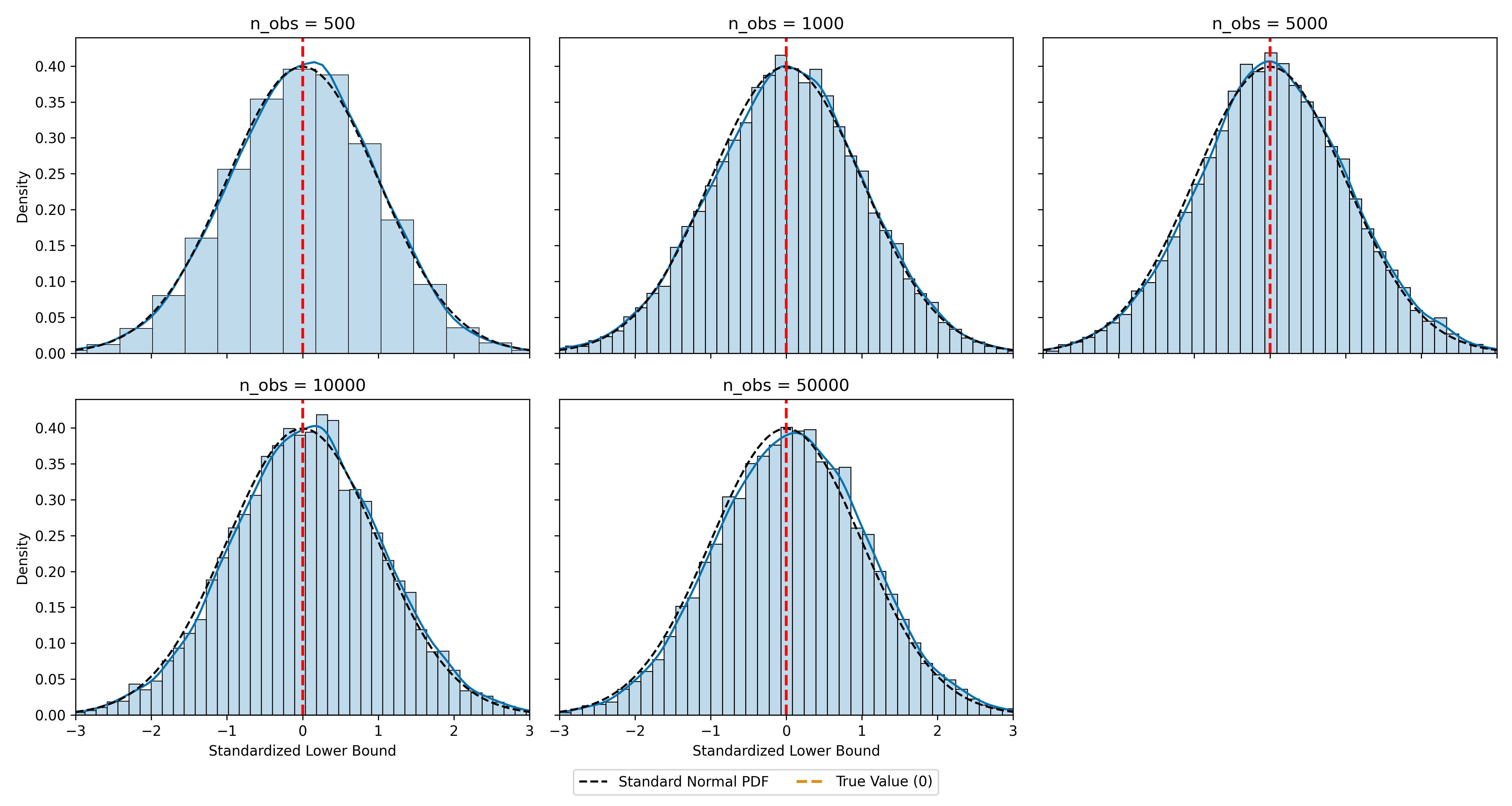} \\
    ($ii$)\\
    \includegraphics[width=0.6\linewidth]{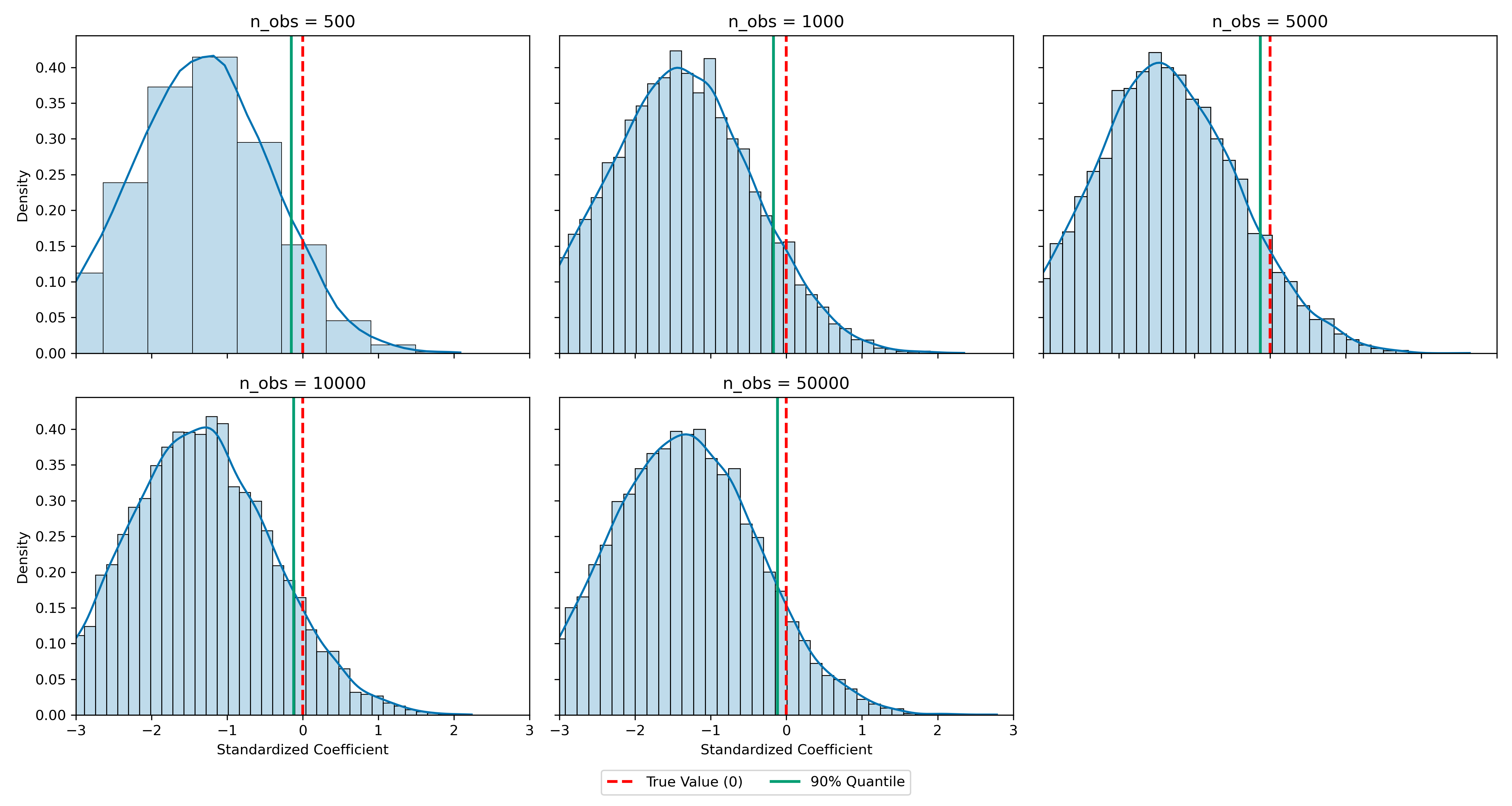} \\
    \caption{ Histograms of ($i$) $\widetilde{\hat{\theta}}_{-,r}=\frac{\hat{\theta}_{-,r} - \theta_0}{\sqrt{\frac{1}{R}\sum_1^R (\hat{\theta}_{-,r} - \overline{\hat{\theta}_-})^2}}$ and ($ii$)  $\widetilde{\hat{\ell}}_{-,r}=\frac{\hat{\ell}_{-,r} - \theta_0}{\sqrt{\frac{1}{R}\sum_1^R (\hat{\ell}_{-,r} - \overline{\hat{\ell}_{-,r}})^2}}$. Results from $R=10,000$ simulation repetitions.}
    \label{fig:histograms}
\end{figure}
\begin{figure}[t]
    \centering
   \includegraphics[width=0.7\linewidth]{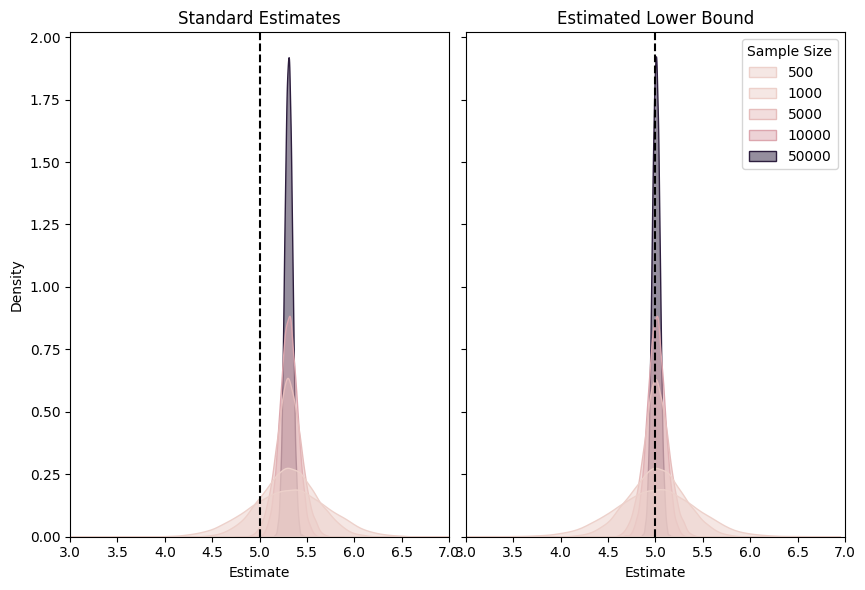}
    \caption{Density plots for DML estimate, $\hat{\theta}_s$, and lower bound, $\hat{\theta}_-$, based on $10,000$ simulation repetitions.}
    \label{fig:density}
\end{figure}

To get more insight on the distribution of the lower bounds for the point estimate and the lower $90\%$ confidence limit, we provide histograms of their standardized versions in Figure \ref{fig:histograms}.  The histograms show that the lower bounds  $\hat{\theta}_-$ (Panel ($i$)) and $\hat{\ell}_-$ (Panel ($ii$)) are normally distributed with $\hat{\theta}_-$ being close to the true ATT, on average. 
The distribution of the lower bias bounds is rather dispersed in small data settings, with the approximation of a standard normal distribution becoming better with larger sample sizes. As indicated by the solid green vertical line, the $90th$ percentile of the empirical distribution of $\hat{\ell}_-$ is close to the true ATT (red dashed line), which provides evidence on the close-to-nominal level empirical coverage of the sensitivity confidence bounds with decreasing estimation uncertainty. Figure \ref{fig:density} illustrates the empirical distribution of the DML estimate (short model), $\hat{\theta}_s$, and the lower bound, $\hat{\theta}_-$, for settings with increasing sample size in a density plot. The figure illustrates that with larger sample size, the lower bias bound (right panel) concentrates around the true value, whereas the DML estimate exhibits a bias irrespective of the reduced estimation uncertainty. 

\begin{figure}[t]
    \centering
   \includegraphics[width=0.7\linewidth]{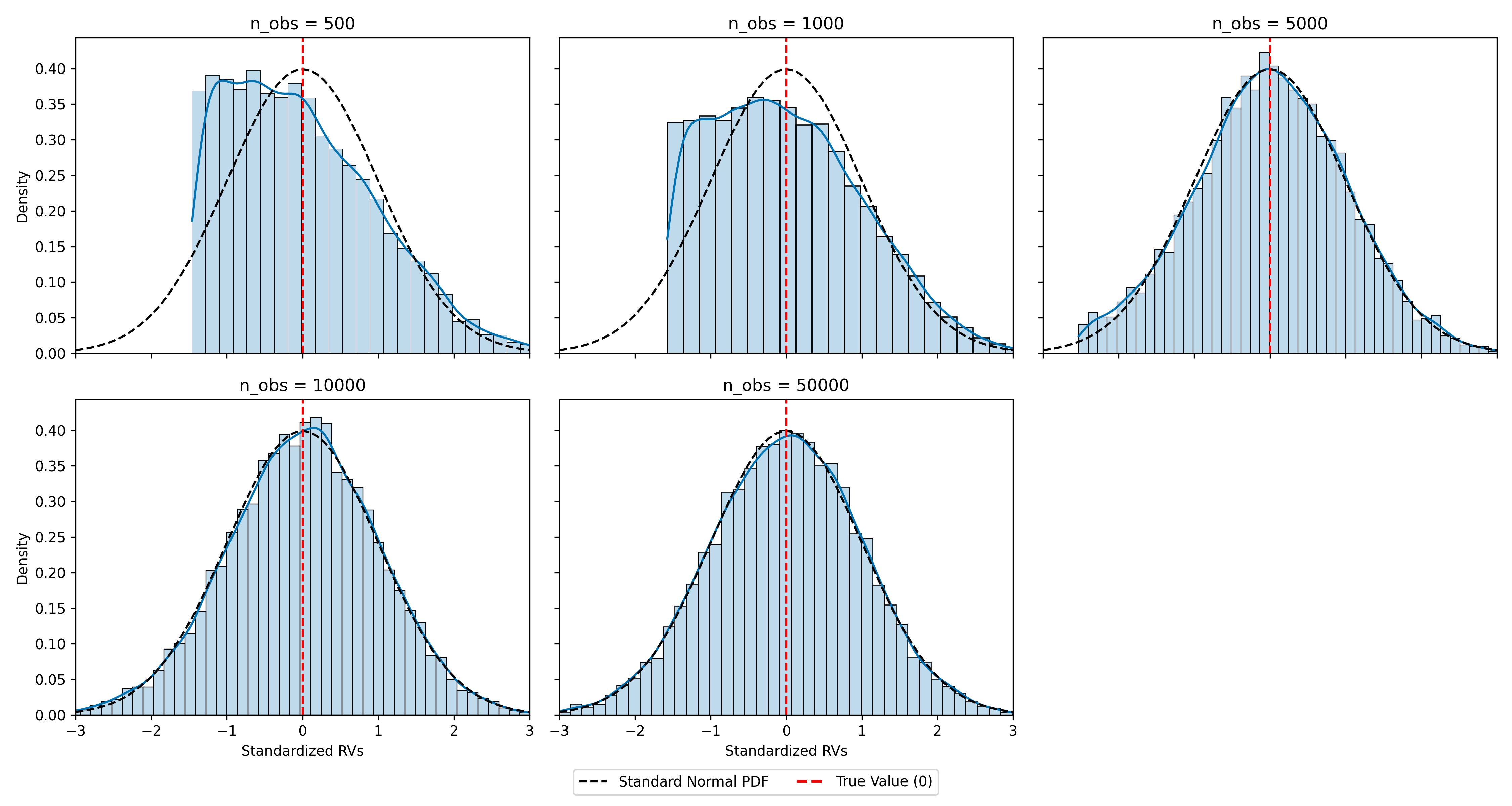}
    \caption{Histograms of standardized robustness values $RV_{\theta=5}$ from $R=10,000$ simulation repetitions.}
    \label{fig:rvs}
\end{figure}

Figure \ref{fig:rvs} shows the histogram of the standardized robustness values. The results for the small sample settings with $n=500$ and $n=1000$ show that the estimation of the $RV$ might be complicated by numerical instabilities, as it is often estimated to be $0$. With moderate and larger samples, the $RV$ approaches its nominal value and exhibits an empirical distribution that is similar to the standard normal distribution. More results can be found in Appendix \ref{app:sim}.

\section{Empirical Application} \label{application}

\subsection{Sensitivity Analysis for the LaLonde Data} 
\label{application:lalonde}

We apply the framework for sensitivity analysis to the famous LaLonde data. In his influential study, \cite{lalonde1986} evaluated the causal effect of participation in the National Supported Work Demonstration (NSW) program on earnings combining two different data sources. First, he used data from a field experiment, where individuals were randomly assigned to participate in a job training. Second, he constructs additional data sets from the Current Population Survey (CPS)\footnote{We follow \cite{smith2005} and focus on using the CPS samples.} and the Panel Study of Income Dynamics (PSID). The idea of using these additional data sets was to mimic an observational study, for which the results could be compared to those of an experimental evaluation. The so-called Lalonde critique casts doubts on the validity of (at that time state-of-the-art) observational causal inference techniques and sparked an intense debate on the use of observational data in the economics and econometrics literature. Important studies include \cite{dehejia1999causal}, \cite{dehejia2002propensity} (henceforth DW), \cite{heckman1997matching}, and \cite{smith2005}, among others. A comprehensive survey and reevaluation has recently been provided by \cite{imbens2024lalonde}, which includes state-of-the-art causal estimators and new data sets. \cite{imbens2024lalonde} also perform sensitivity analysis using the framework of \cite{cinelli2020making} for linear regression under the unconfoundedness assumption.

We perform sensitivity analysis on the ATT in the $2\times 2$ DiD setting building on the previous work by \cite{smith2005}, which was also evaluated using the doubly robust ATT estimator in \cite{sant2020doubly}. Also \cite{huber2024joint} use the Lalonde-PSID sample for their joint test for unconfoundedness and parallel trends.

\cite{smith2005} employ a matching DiD estimator to analyze three different data sets: (1) The original Lalonde-CPS sample, (2) a modified version of the Lalonde-CPS data used in DW, and (3) a refined version of the DW data that focuses on individuals from an early phase of the field experiment. The appealing feature of the LaLonde data is the availability of an experimental benchmark for observational causal estimates. This makes it possible to run two different types of a quasi-observational causal evaluation: First, following \cite{lalonde1986} and DW, the individuals that were actually assigned to participate in the job training are considerd as a treatment group and the control group is composed from the CPS data. Second, it is possible to evaluate a placebo effect: Those individuals who were not assigned to the treatment in the experiment are considered as a pseudo-treated group, for which the ATT is evaluated against the control group from the CPS data. The placebo analysis is useful to quantify the so-called ``evaluation bias'' \cite[P. 320]{smith2005} that originates from systematic differences in the experimental and observational samples. Because the pseudo-treated group did not actually receive the treatment, a nonzero and possibly significant ATT estimate only reflects selection into the experimental sample ($\theta_0=0$ in this case). In their results, \cite{smith2005} and \cite{sant2020doubly} report the evaluation bias as $\frac{\hat{\theta}}{\hat{\theta}_\text{Exp}}$, where $\hat{\theta}_\text{Exp}$ is the ATT estimate from the original experimental evaluation in \cite{lalonde1986}.\footnote{A bias of $1$ would indicate that a reported ATT estimate is to $100\%$ reflecting a bias due to selection into the experimental sample.} 

\begin{table}[t]
\centering
\begin{tabular}{lccc}
\toprule
  Samples & LaLonde  & DW  & Early RA  \\
 ATT, $\theta_0$ & $0$ & $0$ & $0$ \\ 
 Exp. Benchmark & $886$ & $1,794$ & $2,748$ \\ 
$\#$ Treated (Placebo) & $425$ & $260$ & $142$\\
 \midrule
\multicolumn{4}{c}{\textit{(1) Point estimation}} \\ \\[-1.2ex]
 Spec. $g()$ & ADW & ADW & ADW \\
 Spec. $m()$ & DW & DW & DW \\
 Learner $g()$ & Ridge & Lin. Reg. & Lin. Reg. \\
 Learner $m()$ & Ridge & Log. Reg. & Log. Reg. \\ \\
 $\hat{\theta}$ & -692 & 301  & -326 \\
Std. err. & 414 & 487 & 592 \\
Eval. bias ($\%$) & -78 & 17 & -12 \\ \midrule
\multicolumn{4}{c}{\textit{(2) Sensitivity Analysis}} \\  \\[-1.2ex]
RV ($\%$) & 0.008 & 0.003 & 0.003 \\
RV$_\alpha$ ($\%$) & 0.000 & 0.000 &  0.000 \\
\bottomrule
\end{tabular}
\caption{\footnotesize DML DiD estimation results for placebo analysis for different data sets considered in \cite{sant2020doubly} and \cite{smith2005}. Size of control group: $15,992$. DW: Model specification as suggested in DW; ADW: Augmented DW specification of outcome difference regression and propensity score in \cite{sant2020doubly}, Lin.Reg = Linear regression, Log. Reg. = Logistic regression; More details provided in Appendix \ref{app:results}.}
\label{tab:lalonde}
\end{table}

The rationale for the placebo analysis in \cite{smith2005} is that the previously reported observational estimates in DW are not only reflecting the causal effect, which is estimated according to their flexibly specified propensity score matching estimators, but also affected by such sample selection bias. To account for systematic differences in the experimental and observational samples related to the measurement of the outcome variable and accounting for local labor market characteristics, \cite{smith2005} suggest to use matching difference-in-differences estimators. Accordingly, \cite{smith2005} conclude that DiD estimators are better able to adjust for these time-invariant factors.

\begin{figure}[t]
\centering
 \includegraphics[width=0.8\textwidth]{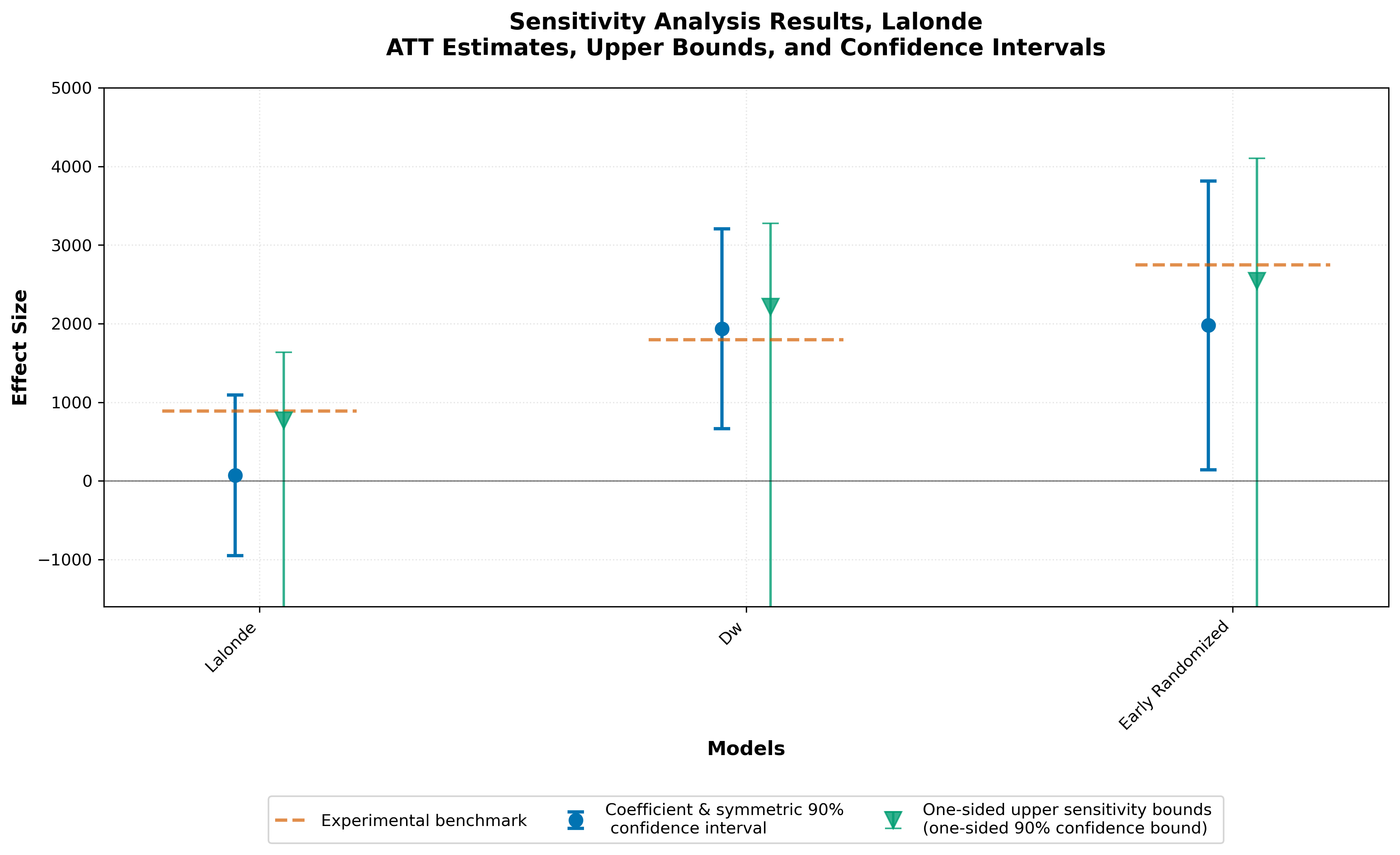}
    \caption{ATT estimates and two-sided $90\%$-confidence levels (blue), experimental benchmarks from \cite{lalonde1986} (dahsed lines), one-sided bias bounds for the ATT with one-sided upper $90\%$-confidence bound for LaLonde data sets (with actually treated group). The bias bounds are evaluated at the robustness values from the placebo analysis presented in Table \ref{tab:lalonde}.}
    \label{fig:lalonde_sensitivity}
\end{figure}

Moreover, \cite{smith2005} emphasize that the functional form specification for the propensity score plays an important role to replicate results in DW, which is in line with the results in \cite{sant2020doubly}. \cite{sant2020doubly} use linear and logistic regression based on manually constructed specifications including polynomial and interaction terms as motivated by DW. Table \ref{tab:lalonde} show the point estimates for the DML DiD ATT in the placebo analysis according to the preferred choices for modeling the nuisance components. We chose the learner specification that performed best in terms of the predictive performance for the nuisance functions $g()$ and $m()$. To address overlap issues, which are known to be a major challenge in the LaLonde data evaluation, we calibrated the learners using isotonic regression \citep{van2024doubly, van2025automatic, klaassen2025calibration, ballinari2024calibrating}. More results, including those from other learner choices, are available in Appendix \ref{app:results}. The DML DiD estimators and the evaluation bias in Panel (1) of Table \ref{tab:lalonde} are all in the range of the results reported in \cite{sant2020doubly} (Table 3), with slightly reduced standard errors. 
Overall, the estimation of the ATT is very variable leading to non-significant estimates, due the number of pseudo-treated individuals being very small as compared to the control group. 
As a consequence the robustness values $RV$ displayed in Panel (2) of Table \ref{tab:lalonde} are very close to zero pointing at non-robust effects. However, we estimate the bias bounds for the ATT in the sample with the actual treated based on the placebo settings, i.e., we set  $C_{\Delta Y}^2$ and $\widetilde{C}_D^2$ to the corresponding robustness values from the placebo analysis. The upper bias bounds for the point estimates and confidence bounds are shown in Figure  \ref{fig:lalonde_sensitivity}. It is possible to see that in the LaLonde CPS and early randomized samples, the upper sensitivity bounds for the ATT are now closer to the experimental benchmark. In all cases, the confidence sensitivity bounds cover the experimental benchmarks.

\subsection{The Impact of Minimum Wages on Firm Profitability} \label{sec:draca}

As a second empirical example, we apply sensitivity analysis to a study by \cite{draca2011minimum} who evaluate the causal effect of the introduction of the national minimum wage (NWM) in the UK on firm profitability in 1999. Unlike the original study, we only consider the case of a balanced panel, i.e., we drop observations that are not observed during the entire study period from 1994 to 2002. The unit of observation $i$ at time $t$ is a firm. In contrast to the original study with data of up to $771$ firms, our balanced panel data set contains $337$ with $57$ treated firms. \cite{draca2011minimum} define the treatment based on pre-treatment wages: Firms that are expected not to be affected or less affected according to average wages in the time before the NWM introduction are categorized as untreated. Firms with lower average wages prior to the introduction of the NWM are considered as treated. In line with the original study, we consider the effect of the NWM introduction on two outcome variables, the log average wage at the firm (\texttt{ln\_avwage}) and the firm's profit margin defined as the profit to sales ratio (\texttt{net\_pcm}). In this section, we focus on the analysis with regard to \texttt{net\_pcm} and provide the results on average wages in Appendix \ref{app:results}. The original study employs two-way fixed effects regression. In our analysis, we focus on the previously presented DML DiD estimator in the multi-period setting. Due to the different sample compositions, our results deviate slightly from the original study. 
\begin{figure}[t]
    \centering
    \includegraphics[width=0.8\linewidth]{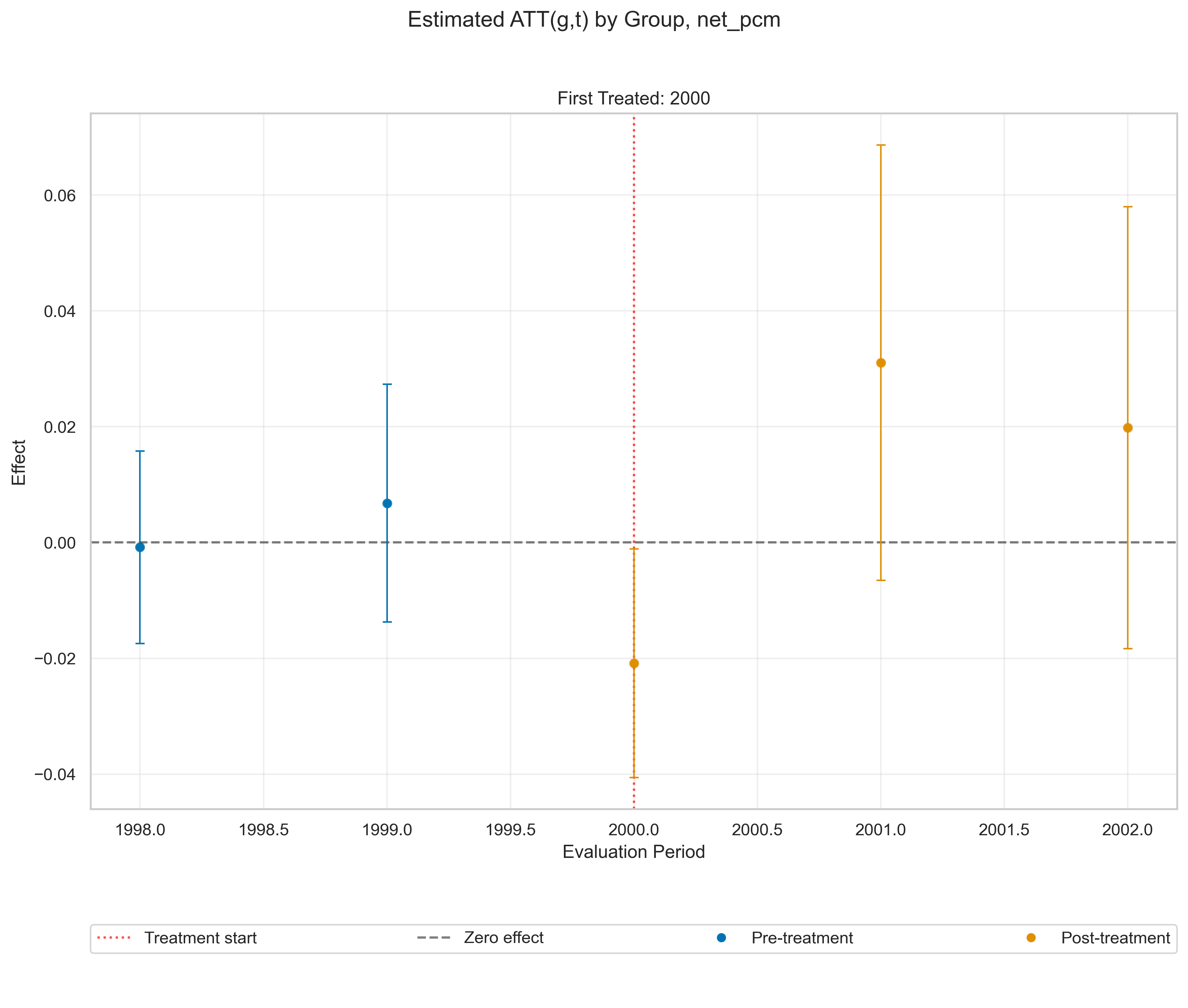}
    \caption{$ATT(\g,t)$ estimates and $95\%$ (pointwise) confidence intervals associated with the 1999-2000 NWM introduction on the net profit margin.}
    \label{fig:attgt_draca}
\end{figure}
In our replication of \cite{draca2011minimum}, we base identification on the conditional parallel trends assumption, which we assume to hold after accounting for pre-treatment information on a firm's industry (2-digit industry classification, \texttt{sic2}), the government office region of workplace (\texttt{gorwk}), the share of part-time workers (\texttt{ptwk}), the share of female workers (\texttt{female}), and the share of union members (\texttt{unionmem}) by the three-digit industry classification. 
The latter three of these variables are time variant whereas there is no variation in industry and region information over time. For the time-varying variables, we condition on the pre-treatment level. 
Figure \ref{fig:attgt_draca} illustrate the $ATT(\g,t)$ estimates as obtained from double machine learning using a random forest learner for the propensity score and the outcome difference regression. We estimate an effect of the NWM introduction on the net profit margin in the first post-treatment period of $-0.021$ ($95\%$ confidence interval $[-0.041,-0.001]$), which is in line with the results in \cite{draca2011minimum}.

\begin{figure}[t]
    \centering
    \includegraphics[width=0.9\linewidth]{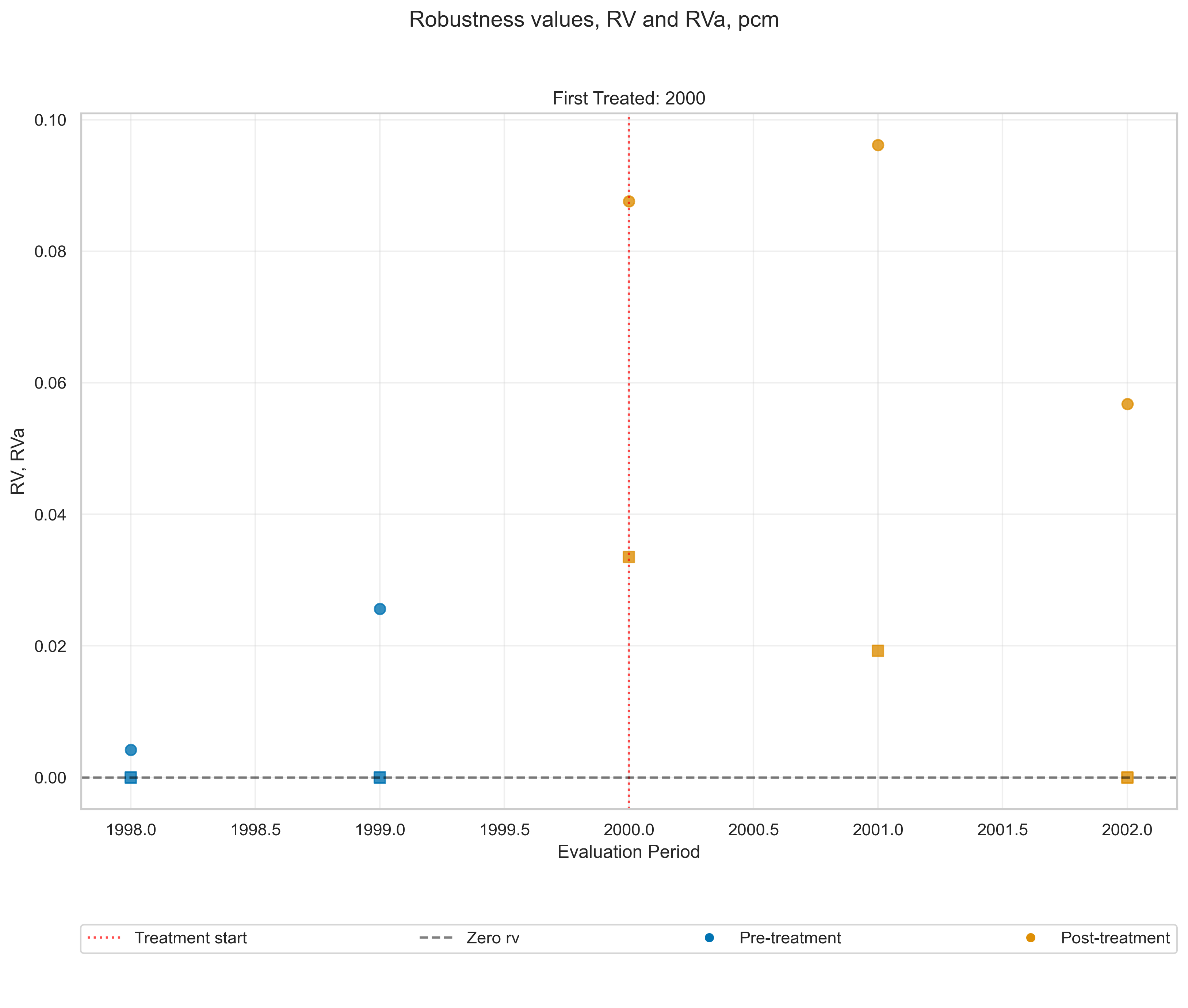}
    \caption{Robustness values $RV$ and $RV_{a=0.1}$ for the $ATT(\g,t)$ parameters evaluating the 1999-2000 NWM introduction the net profit margin.}
    \label{fig:rvs_draca}
\end{figure}

We apply our sensitivity approach to the $ATT(\g,t)$ parameters and show the resulting robustness values in Figure \ref{fig:rvs_draca}. 
The $RV$s for the post-treatment $ATT(\g,t)$ estimates range between $6.5\%$ for the third post-treatment period to $9.6\%$ in the second post-treatment period. To set these values into context, we can exploit the information from pre-testing, for which we expect non-significant effects and, thus, small $RV$ values under valid conditional parallel trends (cf. Section \ref{sec:scenarios}). The effect estimates for the pre-treatment periods are relatively close to zero and not significant. The maximum robustness value from pre-testing is $2.56\%$ as obtained for the last pre-treatment period. 
Compared to the post-treatment $RV$ values , the pre-testing $RV$ is relatively small. Note that the pre-testing coefficient also has a different sign than the post-treatment $ATT(\g,t)$ estimate, which might, hence, result in a rather conservative parallel trend violation scenario.

An appealing feature of the study by \cite{draca2011minimum} is the availability of a rich set of pre-treatment confounding variables, which can be exploited for benchmarking exercises and, thus, inform plausible parallel trend violation scenarios. Following the general benchmarking procedure described in Appendix F of \cite{chernozhukov2022long}, we can compute gain statistics from leaving out covariates. The corresponding values for the sensitivity parameters are presented in Table \ref{tab:bench_npm}. In many cases, we find that omitting the covariates, as for example the information on industry classification, has considerably more explanatory power for the treatment status than for the difference in outcomes. This is in line with the general intuition that  information on a firm's industry is likely related to the average wage level (prior to the NMW introduction), which is used to define the treatment status in \cite{draca2011minimum}.

The previous sensitivity exercises have resulted in a set of parallel trend violation scenarios. As a next step, we consider the critical parallel trend violation that would lead to a substantial change in the causal results, i.e., reduce the (negative) effect estimate of the NWM introduction to zero. In the following, we focus on the sensitivity bounds for the $ATT(\g,t)$ in the first post-treatment period. For this parameter, we obtain robustness values $RV=8.76\%$ and $RV_{a=0.1}=1.81$. Hence, an unobserved pre-treatment confounder $A$, which could explain $8.76\%$ of the residual variation in the conditional expected outcome difference and lower the odds ratio for the treatment status according as explained by the oracle model by $8.76\%$ would be required in order to set the effect to zero. 

\begin{figure}[t]
    \centering
    \includegraphics[width=\linewidth]{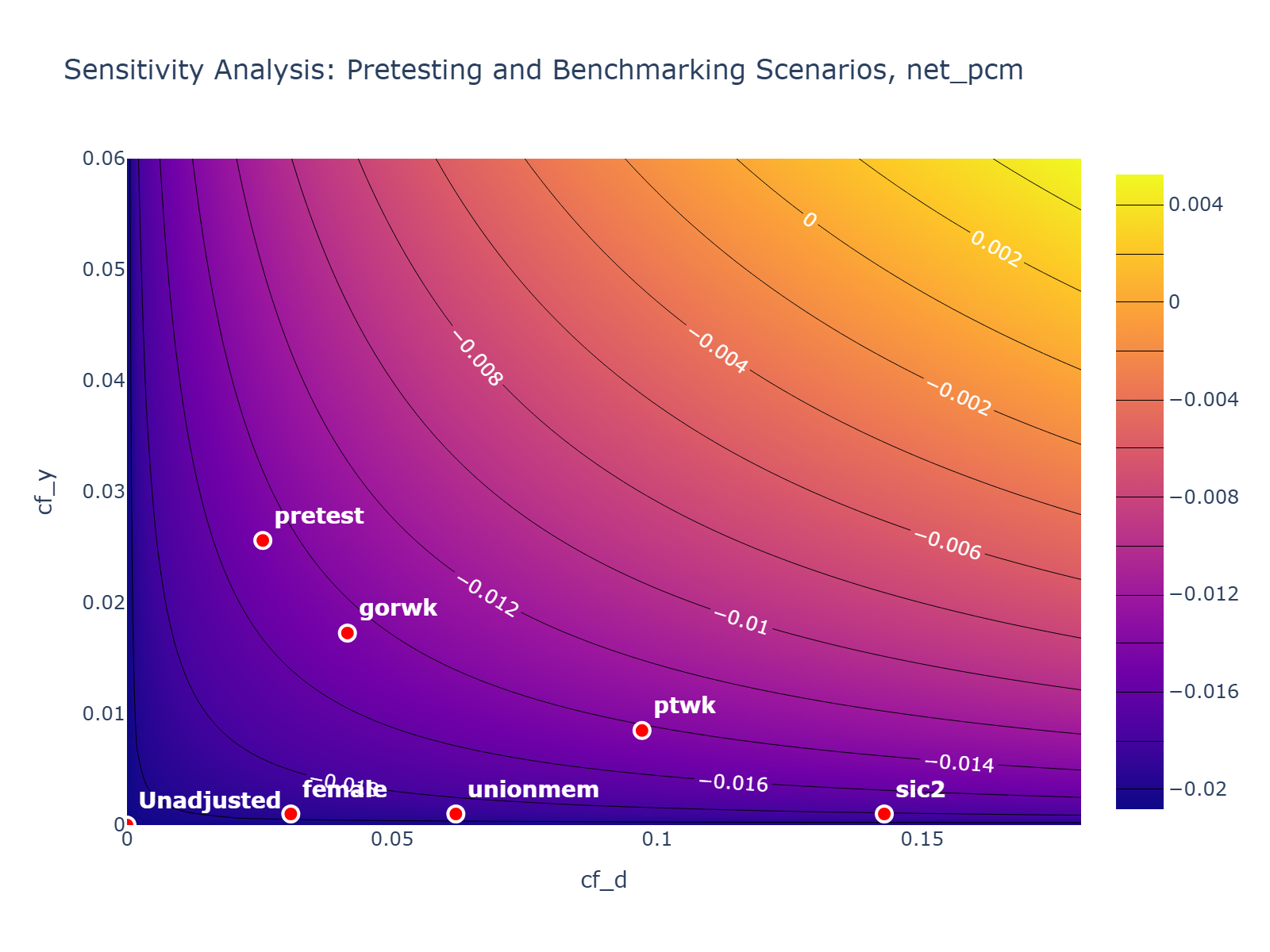}
    \caption{Contour plot illustrating the upper bound of the $ATT(\g,t)$ estimate for the first post-treatment period in different scenarios of parallel trends violations. The indicated scenarios illustrate the values from pretesting and selected benchmarking settings, with $\rho=1$ being enforced.}
    \label{fig:contour_draca}
\end{figure}

As a next step, we can estimate the upper sensitivity bounds for the $ATT(\g,t)$ parameter in the first post-treatment period. To do so, we use an additional visualization of the upper bound for the causal parameter in the different parallel trend violation scenarios through a contour plot, presented in Figure \ref{fig:contour_draca}. Note that we enforced $\lvert \rho \lvert=1$ for the indicated scenarios in the contour plot to maintain the comparability of the different scenarios.\footnote{$\rho$ operates as a scaling factor in the bias formula in Equation \eqref{eq:bias}.} As a consequence, the evaluated settings are worst-case scenarios with a maximum correlation of the confounding variation in the treatment assignment and outcome difference. The contour plot shows that the upper bound for the $ATT(\g,t)$ parameter in the first post-treatment period is negative in all scenarios considered. A parallel trend violation corresponding to the strongest pre-treatment violation would result in a reduction of the causal effect (in absolute values) to $-0.015$. The strongest (conservative) benchmarking scenario corresponds to leaving out the share of part-time workers by the three-digit industry classification. A parallel trend violation that is comparably strong in terms of the explanatory power for the treatment assignment and the considered outcome difference would induce an adjustment of the $ATT(\g,t)$ parameter to $-0.014$.

\begin{table}[t]
    \centering
\begin{tabular}{l lrrr}
\toprule
 $Y$ & Scenario & $\widetilde{C}_D^2$ & $C_{\Delta Y}^2$ & $\lvert \rho\lvert$ \\
\midrule
net\_pcm & \textit{Pretest} &  0.0256 & 0.0256 & 1.0000 \\
& \textit{Benchmark}  & & & \\
& - ptwk &  0.0972 &  0.0085 &  0.1299  \\
& - sic2 & 0.1429 & 0.0010 & 1.0000 \\
& - unionmen & 0.0620 & 0.0010 & 1.0000 \\
& - female & 0.0309 & 0.0010 & 1.0000\\
& - gorwk & 0.0416 & 0.0173 & 0.1171  \\
\bottomrule
\end{tabular}
    \caption{PT violation scenarios based on pre-testing and benchmarking pre-treatment covariates. We enforce a minimum value of $0.0010$ for benchmarking scenarios.}
    \label{tab:bench_npm}
\end{table}

\begin{figure}[t]
    \centering
    \includegraphics[width=0.7\linewidth]{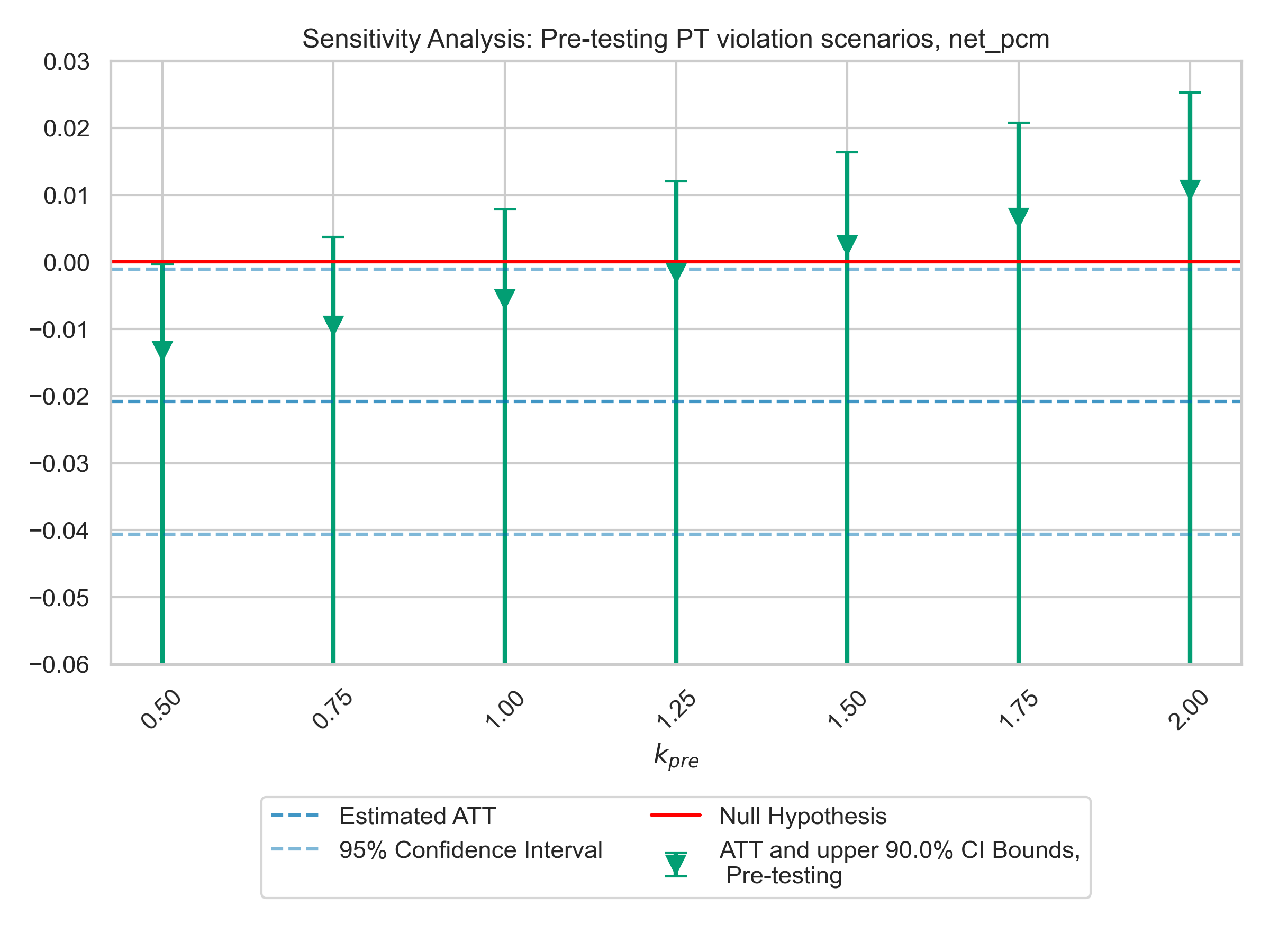}
    \caption{$ATT(\g,t)$ for the first post-treatment period with one-sided $90\%$ confidence bounds according to different PT violation scenarios based on pre-testing, dependent variable \textit{net\_pcm}.}
    \label{fig:bounds_draca_pcm}
\end{figure}

Moreover, it is possible to vary the strength in terms of a $k$-fold multiple of the pretesting scenario sensitivity parameters, which would be similar to the type of results commonly reported in applications of the approach by \cite{rambachan2023}. Figure \ref{fig:bounds_draca_pcm} presents the upper bound of the $ATT(\g,t)$ parameter in the first post-treatment period and the one-sided upper confidence bound at a nominal level of $90\%$. The figure illustrates that the reported causal effect in the first post-treatment would become non-significant if the conditional parallel trend violation was comparable to the pre-testing scenario.

The previous sensitivity results point at a rather robust effect in the considered scenarios. Of course, the overall conclusion from our empirical analyses have to be set into the context of the study. \cite{draca2011minimum} study a country-wide introduction of a minimum wage, which is a policy affecting all companies operating in the UK. Differences in the probability to be affected by the NMW introduction and the change in the profitability over the considered time horizon might be related to firm characteristics, such as industry and characteristics of the workers. We find that benchmarking against observed variables in the data point at a rather robust effect. However, considering estimation uncertainty, the results in Figure \ref{fig:bounds_draca_pcm} show that the effect quickly becomes non-significant in our pre-testing sensitivity exercises. 

 \pagebreak

\section{Conclusion and Outlook} \label{conclusion}

Our study contributes to the existing and quickly growing literature on difference-in-differences, causal machine learning and sensitivity analysis.  
In the previous section, we presented a new approach to sensitivity analysis with regard to violations of the conditional parallel trend assumption in common difference-in-differences models. In many empirical applications, researchers are concerned with potential violations of the parallel trend assumptions and a corresponding bias of the causal parameter estimate. To assess the robustness of causal estimation results due to such violations, researchers can use our suggested sensitivity approach and obtain asymptotic bounds for the point estimates and confidence intervals for the target parameters. Specific violation scenarios in terms of a set of values for the sensitivity parameters in the asymptotic bias formula can be based on pre-testing evidence, benchmarking analyses, standard reporting statistics and domain expertise. In addition to the theoretical results on Riesz representation in the canonical and multi-period DiD setting with staggered adoption, we provide new evidence on the validity of Riesz-representation-based sensitivity analysis in a simulation study. Moreover, we provide an open source implementation of our approach in \texttt{DoubleML} for Python and we demonstrate the application of DiD sensitivity analysis in two empirical examples. 

There are various extensions worth to explore in future research. A natural extension of the current approach is to aggregations of the $ATT(\g,t)$ parameters in the multi-period difference-in-differences setting as considered in \cite{callaway2021difference}, For example, often the treatment effect is evaluated relative to the treatment receipt in event studies. Riesz-representation based sensitivity analysis for aggregated effects is non-trivial and, hence, left for future research. Further extensions might be interesting for recently considered difference-in-differences models such as models with continuous treatments. Finally, empirical application of Riesz-representation-based sensitivity analysis is important to address empirical challenges of these new techniques for causal analysis.

\pagebreak
\bibliographystyle{plainnat} 
\bibliography{bibliography} 

\clearpage

\appendix
\section{Appendix} \label{appendix}

\subsection{Derivation of the Riesz Representer for Difference-in-Differences Estimators} \label{app:rr_binary}
Under Assumptions \ref{assumption::parallel_trends} and \ref{assumption::overlap}, 
the ATT is identified by
\begin{align*}
    \theta_0 
    &= \E\left[Y_{1}(1) - Y_{1}(0)|D=1\right] \\ 
    &= \E\left[Y_{1}(1) - Y_{0}(0)|D=1\right] - \E\left[Y_{1}(0) - Y_{0}(0)|D=1\right]\\
    &= \E\left[\Delta Y|D=1\right] - \E\left[\E\left[Y_{1}(0) - Y_{0}(0)|D=1, X, A\right]|D=1\right]\\ 
    &= \E\left[\Delta Y|D=1\right] - \E\left[\E\left[Y_{1}(0) - Y_{0}(0)|D=0, X, A\right]|D=1\right]\\
    &= \E\left[\Delta Y|D=1\right] - \E\big[\underbrace{\E\left[\Delta Y|D=0, X,A\right]}_{=:g(0,X,A)}|D=1\big]\\
    &= \E\left[\Delta Y - g(0,X,A)|D=1\right]\\
    &= \E\left[\frac{D}{p}\big(\Delta Y - g(0,X,A)\big)\right]\\
    &= \E\bigg[\underbrace{\frac{D}{p}\big(g(1,X,A) - g(0,X,A)\big)}_{=:\mathcal{M}(W,g)}\bigg]
\end{align*}
where $W=(\Delta Y, D, X,A)$. Next, we derive the Riesz representation. We have
\begin{align*}
    \alpha(W) = \frac{D}{p} - \frac{(1-D)}{p}\frac{m(X,A)}{1-m(X,A)}
\end{align*}
since 
\begin{align*}
    \E\left[g(W)\alpha(W)\right] 
    &= \E\left[g(W)\left(\frac{D}{p} - \frac{(1-D)}{p}\frac{m(X,A)}{1-m(X,A)}\right)\right]\\
    &= \E\left[\frac{D}{p}\Delta Y - \frac{(1-D)}{p}\frac{m(X,A)}{1-m(X,A)}g(0,X,A)\right]\\
    &= \E\left[\frac{D}{p} (\Delta Y - g(0,X,A))\right] + \E\left[g(0,X,A)\left(\frac{D}{p} - \frac{(1-D)}{p}\frac{m(X,A)}{1-m(X,A)}\right)\right]\\
    &= \E\left[\frac{D}{p} (g(1,X,A) - g(0,X,A))\right]= \E[\mathcal{M}(W,g)]
\end{align*}
using
\begin{align*}
&\E\left[g(0,X,A)\left(\frac{D}{p} - \frac{(1-D)}{p}\frac{m(X,A)}{1-m(X,A)}\right)\right]\\
&\E\left[\E\left[\Delta Y|D=0, X,A\right]\left(\frac{D}{p} - \frac{(1-D)}{p}\frac{m(X,A)}{1-m(,A)}\right)\right]\\
    =\  &\E\left[\E\left[\Delta Y|D=0, X,A\right]\E\left[\left(\frac{D}{p} - \frac{(1-D)}{p}\frac{m(X,A)}{1-m(X,A)}\right)|X,A\right]\right]\\
    =\ &0.
\end{align*}

\subsection{Derivation of the bias formula}

With the same argument used in \ref{app:rr_binary}, one can show that
\begin{align*}
    \E\left[g_s(W)\alpha_s(W^s)\right] &= \E[\mathcal{M}(W^s,g_s)]
\end{align*}
such that
\begin{align*}
    \theta_0 - \theta_s &= \E\left[\left(g(W)-g_s(W^s)\right)(\alpha(W)-\alpha_s(W^s)\right].
\end{align*}
Theorem 2 in \cite{chernozhukov2022long} implies that
\begin{align*}
    |\theta_0 - \theta_s|^2=\rho^2B^2\le B^2
\end{align*}
where $B^2:=\E[(g(W)-g_s(W^s))^2]\E[(\alpha(W)-\alpha_s(W^s))^2]$ and 
$$\rho^2:=Cor^2(g(W)-g_s(W^s),\alpha(W)-\alpha_s(W^s))\le 1.$$
Corollary 2 in \cite{chernozhukov2022long} concludes that
\begin{align*}
|\theta_0 - \theta_s|^2 = C_{\Delta Y}^2 C_D^2 S^2,
\end{align*}
with $S^2 = \E[(\Delta Y-g_s(W^s))^2 \E[\alpha_s^2(W^s)]$. The sensitivity parameters are given by
\begin{align*}
 C_{\Delta Y}^2 &= \frac{\E[(g-g_s)^2]}{\E[(\Delta Y-g_s)^2]} =  \frac{\V\big(\E[\Delta Y|D, X, A]\big) - \V\big(\E[\Delta Y|D, X]\big)}{\V\big(\Delta Y\big) - \V\big(\E[\Delta Y|D, X]\big)} \\
 C_D^2 &= \frac{\E[\alpha^2] - \E[\alpha_s^2]}{\E[\alpha^2_s]} = \frac{\E\left[\frac{m(X, A)}{1-m(X,A)}\right] - \E\left[\frac{m(X)}{1-m(X)}\right]}{\E\left[\frac{m(X)}{1-m(X)}\right]},
\end{align*}
where we used that
\begin{align*}
    \E[\alpha(W)^2] 
    &= \E\left[\left(\frac{D}{p} - \frac{1-D}{p}\frac{m(X)}{1-m(X)}\right)^2\right]\\
    &= p^{-2}\E\left[D + (1-D)\frac{m(X)^2}{(1-m(X))^2}\right]\\
    &= p^{-2}\E\left[m(X) + \frac{m(X)^2}{1-m(X)}\right]\\
    &= p^{-2}\E\left[\frac{m(X)}{1-m(X)}\right].
\end{align*}

\subsection{Sensitivity Parameters for Multi-Period DiD}
In this section, we show that the interpretation of the confounding parameters in the multi-period setting setting is consistent with the 2x2 setting, by considering conditional residual variations.
At first, remark that it holds
\begin{align*}
    \E[\alpha(W)^2] 
    &= \E\left[\left(\left(\frac{G^{\g}}{\E[G^{\g}]} - \frac{m(X,A)(1-G^{\g})}{\E[G^{\g}](1-m(X,A))}\right) \cdot \max(G^{\g}, C^{(\cdot)})\right)^2\right]\\
    &= \E[G^{\g}]^{-2}\E\left[\left(G^{\g} + (1-G^{\g})\frac{m(X,A)^2}{(1-m(X,A))^2}\right)\cdot \max(G^{\g}, C^{(\cdot)})\right]\\
    &= \frac{P(\max(G^{\g}, C^{(\cdot)})=1)}{\E[G^{\g}]^{2}}\E\left[\left(G^{\g} + (1-G^{\g})\frac{m(X,A)^2}{(1-m(X,A))^2}\right) \mid \max(G^{\g}, C^{(\cdot)}) = 1\right]\\
    &= \frac{P(\max(G^{\g}, C^{(\cdot)})=1)}{\E[G^{\g}]^{2}}\E\left[\frac{m(X,A)}{(1-m(X,A))} \mid \max(G^{\g}, C^{(\cdot)}) = 1\right].
\end{align*}
This directly implies
\begin{align*}
     C_D^2 &= \frac{\E[\alpha^2] - \E[\alpha_s^2]}{\E[\alpha^2_s]} = \frac{\E\left[\frac{m(X, A)}{1-m(X,A)}\mid \max(G^{\g}, C^{(\cdot)}) = 1\right] - \E\left[\frac{m(X)}{1-m(X)}\mid \max(G^{\g}, C^{(\cdot)}) = 1\right]}{\E\left[\frac{m(X)}{1-m(X)}\mid \max(G^{\g}, C^{(\cdot)}) = 1\right]}.
\end{align*}
Analogously to \cite{chernozhukov2022long}, let
\begin{align*}
    \widetilde{g}(D,X,A) &:= \E[\Delta Y \mid D, X, A]\\
    \widetilde{g}_s(D,X) &:= \E[\Delta Y \mid D, X]
\end{align*}
denote the respective conditional expectations. As in \cite{chernozhukov2022long}, the bias bound can be expressed using 
\begin{align*}
 \widetilde{C}_{\Delta Y}^2 &:= \frac{\E[(\widetilde{g}-\widetilde{g}_s)^2]}{\E[(\Delta Y-\widetilde{g}_s)^2]} = \frac{\V\big(\E[\Delta Y|D, X, A]\big) - \V\big(\E[\Delta Y|D, X]\big)}{\V\big(\Delta Y\big) - \V\big(\E[\Delta Y|D, X]\big)}
\end{align*}
as measure of confounding strength. This measures the residual variation in $\Delta Y$ explained by $A$, but not only restricted to the current treatment group $G^{\g}$ and control group $C^{(\cdot)}$. This formula does not directly correspond to the interpretation of the 2x2 setting restricted to the treatment and control groups. Further it might be counterintuitive since e.g. additional not yet treated groups could reduce or increase the residual variation but would not actually have any effect on the current bias.
Instead we only fit two parts of the regression function i.e.
\begin{align*}
    g(0,X,A) &:= \mathbb{E}[\Delta Y|X, A, C^{(\cdot)} = 1]\\
    g(1,X,A) &:= \mathbb{E}[\Delta Y|X, A, G^{\g} = 1],
\end{align*}
and correspondingly
\begin{align*}
    g_s(0,X) &:= \mathbb{E}[\Delta Y|X, C^{(\cdot)} = 1]\\
    g_s(1,X) &:= \mathbb{E}[\Delta Y|X, G^{\g} = 1],
\end{align*}
which does not account for variation in $\Delta Y$ outside of treatment and control group. Considering the bias formula of Theorem 2 in \cite{chernozhukov2022long}
\begin{align*}
    \theta_s - \theta 
    &= \E\left[(\widetilde{g} - \widetilde{g}_s)(\alpha - \alpha_s) \right]\\
    &= \E\left[(\widetilde{g} - \widetilde{g}_s)\max(G^{\g}, C^{(\cdot)})(\alpha - \alpha_s) \right]
\end{align*}
due to the indicator $\max(G^{\g}, C^{(\cdot)})$ in $\alpha$ and $\alpha_s$. Using the same decomposition as in Theorem 2 from \cite{chernozhukov2022long}, it follows
\begin{align*}
    |\theta_s - \theta|^2\le\rho^2B^2,
\end{align*}
with 
\begin{align*}
    \rho^2 &:= \Cor^2\left((\widetilde{g} - \widetilde{g}_s)\max(G^{\g}, C^{(\cdot)}),(\alpha - \alpha_s)\right)\\
    B^2 &:= S^2 C_{\Delta Y}^2 C_D^2.
\end{align*}
At first, remark that
\begin{align*}
    \rho^2
    &:= \Cor^2\left((\widetilde{g} - \widetilde{g}_s)\max(G^{\g}, C^{(\cdot)}),(\alpha - \alpha_s)\right) \\
    &= \frac{\left(\E\left[(\widetilde{g} - \widetilde{g}_s)\max(G^{\g}, C^{(\cdot)})(\alpha - \alpha_s)\right]\right)^2}{\E[(\widetilde{g} - \widetilde{g}_s)^2\max(G^{\g}, C^{(\cdot)})]\E[(\alpha - \alpha_s)^2]}\\
    &= \frac{\left(\E\left[(\widetilde{g} - \widetilde{g}_s)(\alpha - \alpha_s)\mid\max(G^{\g}, C^{(\cdot)}) = 1\right]\right)^2}{\E[(\widetilde{g} - \widetilde{g}_s)^2\mid\max(G^{\g}, C^{(\cdot)}) = 1]\E[(\alpha - \alpha_s)^2\mid\max(G^{\g}, C^{(\cdot)}) = 1]}
\end{align*}
Further, it holds
\begin{align*}
    C_{\Delta Y}^2 &:= \frac{\E[(\widetilde{g} - \widetilde{g}_s)^2\max(G^{\g}, C^{(\cdot)})]}{\E[(\Delta Y-\widetilde{g}_s)^2\max(G^{\g}, C^{(\cdot)})]} = \frac{\E[(\widetilde{g} - \widetilde{g}_s)^2\mid \max(G^{\g}, C^{(\cdot)}) = 1]}{\E[(\Delta Y-\widetilde{g}_s)^2\mid \max(G^{\g}, C^{(\cdot)}) = 1]}
\end{align*}
and
\begin{align*}
    S^2 
    &:= \E[(\Delta Y-\widetilde{g}_s)^2\max(G^{\g}, C^{(\cdot)})]\E[\alpha_s^2]\\
    &= P(\max(G^{\g}, C^{(\cdot)}) = 1)^2\E[(\Delta Y-\widetilde{g}_s)^2\mid \max(G^{\g}, C^{(\cdot)}) = 1]\E[\alpha_s^2\mid \max(G^{\g}, C^{(\cdot)}) = 1].
\end{align*}

\subsection{Riesz Representation with In-Sample Normalization}

The orthogonal score function for the $ATT(\g,t)$ parameter with in-sample normalization is given by
\begin{align}\begin{aligned}\widetilde{\psi}(W,\theta, \eta)
:&= -\frac{G^{\mathrm{g}}}{\mathbb{E}_n[G^{\mathrm{g}}]}\theta + \left(\frac{G^{\mathrm{g}}}{\mathbb{E}_n[G^{\mathrm{g}}]} - \frac{\frac{m(X) (1-G^{\mathrm{g}})}{1-m(X)}}{\mathbb{E}_n\left[\frac{m(X) (1-G^{\mathrm{g}})}{1-m(X)}\right]}\right) \left(Y_{t_\text{eval}} - Y_{t_\text{pre}} - g(0,X)\right)\\&= \widetilde{\psi}_a(W; \eta) \theta + \widetilde{\psi}_b(W; \eta)\end{aligned}\end{align}
Based on this score function, the Riesz representation can be derived as
\begin{align}\begin{aligned}m(W,g) &= \big(g(1,X) - g(0,X)\big)\cdot \frac{G^{\mathrm{g}}}{\mathbb{E}[G^{\mathrm{g}}]}\cdot \max(G^{\mathrm{g}}, C^{(\cdot)})\\\alpha(W) &= \left(\frac{G^{\mathrm{g}}}{\mathbb{E}[G^{\mathrm{g}}]} - \frac{\frac{m(X)(1-G^{\mathrm{g}})}{1-m(X)}}{\mathbb{E}\left[\frac{m(X)(1-G^{\mathrm{g}})}{1-m(X)}\right]}\right) \cdot \max(G^{\mathrm{g}}, C^{(\cdot)}).\end{aligned}\end{align}

\clearpage
\section{Additional Results: Simulation Study} \label{app:sim}

\begin{figure}[ht]
    \centering
   \includegraphics[width=0.7\linewidth]{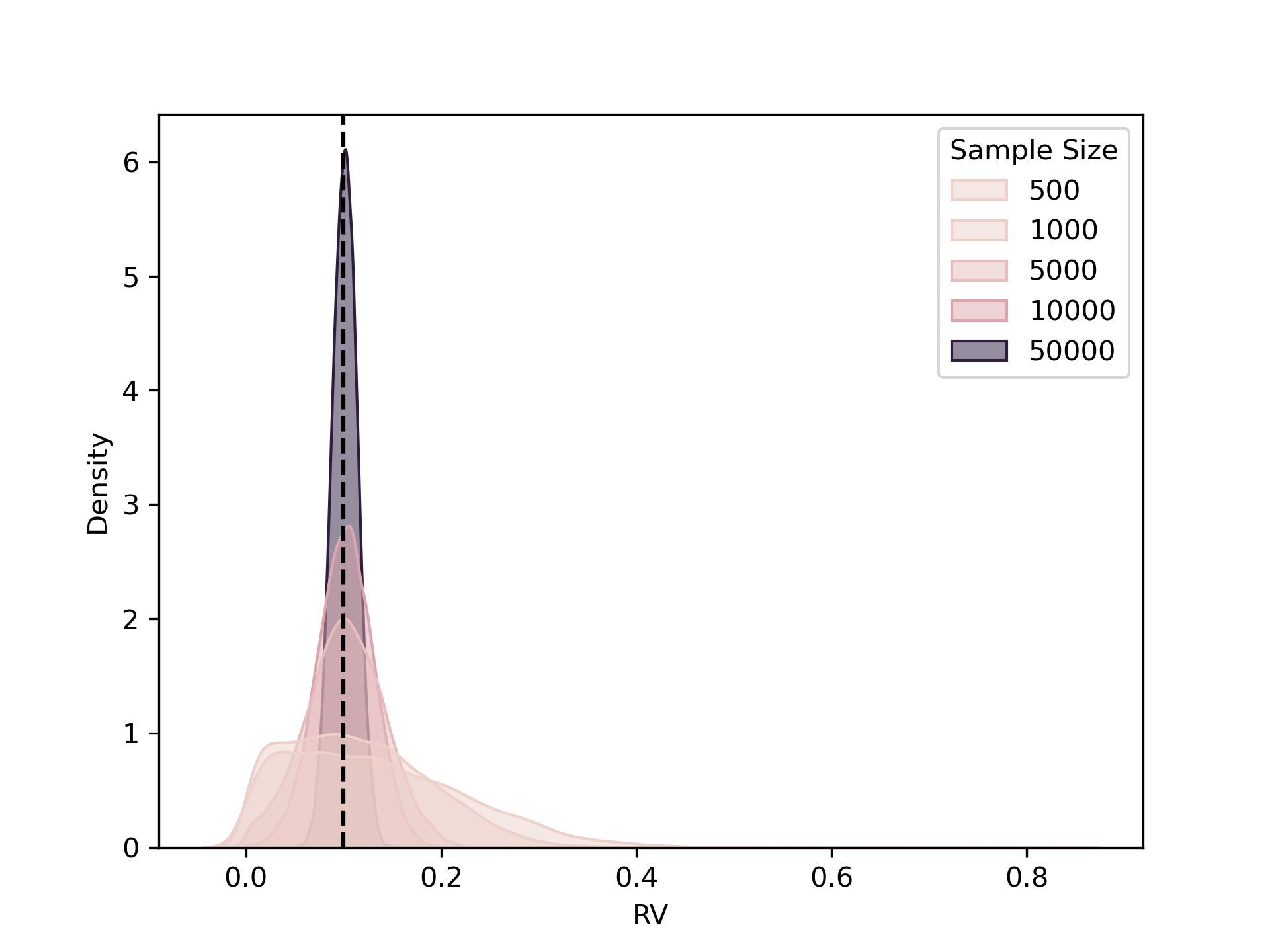}
    \caption{Density plots for robustness value, $RV_{\theta=5}$, based on $10,000$ simulation repetitions.}
    \label{fig:rv_raw}
\end{figure}

\begin{figure}[ht]
    \centering
   \includegraphics[width=0.7\linewidth]{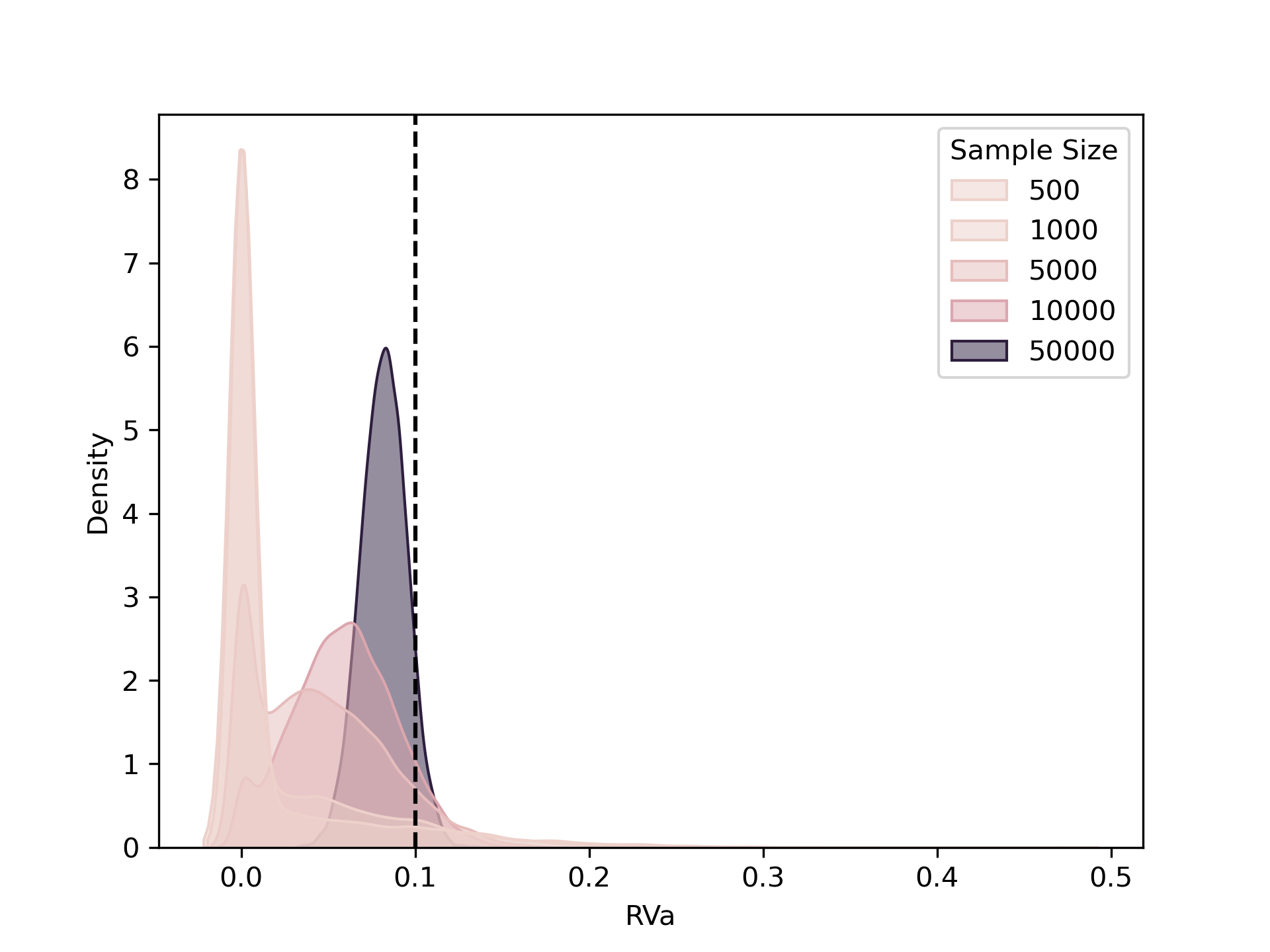}
    \caption{Density plots for robustness value, $RV_{\theta=5,a=0.1}$, based on $10,000$ simulation repetitions.}
    \label{fig:rva_raw}
\end{figure}

\clearpage
\section{Additional Results: Empirical Applications} \label{app:results}

\subsection{Empirical Application on LaLonde Data: Details and Additional Results} \label{app:lalonde}

\begin{figure}
    \centering
    \includegraphics[width=0.85\linewidth]{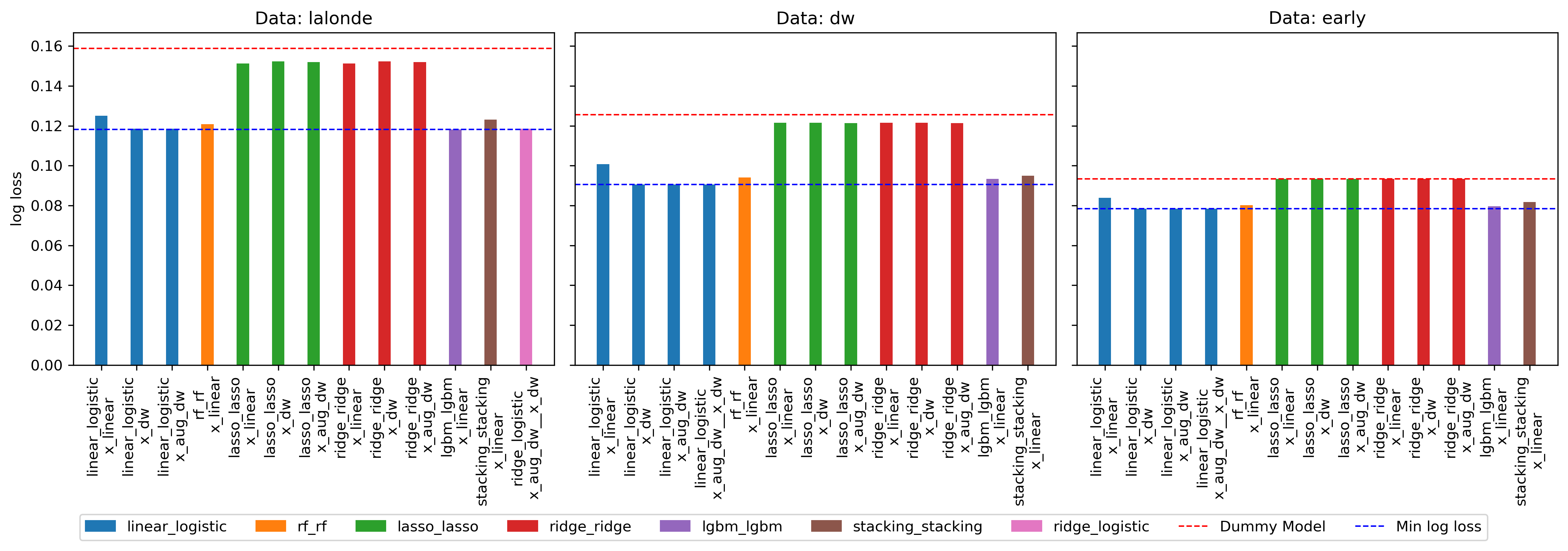}
    \caption{Log loss for different learners and model specifications for the propensity score, evaluated on the out-of-sample predictions from cross-fitting. All learners have been calibrated with isotonic regression. The red line indicates the log loss for a dummy learner, predicting $D=1$ from a binomial distribution with baseline probability $\hat{p}=\frac{1}{n}\sum_i^n d_i$, corresponding to guessing at the unconditional treatment probability. LaLonde data example.}
    \label{fig:log_loss_lalonde}
\end{figure}

\begin{figure}
    \centering
    \includegraphics[width=0.85\linewidth]{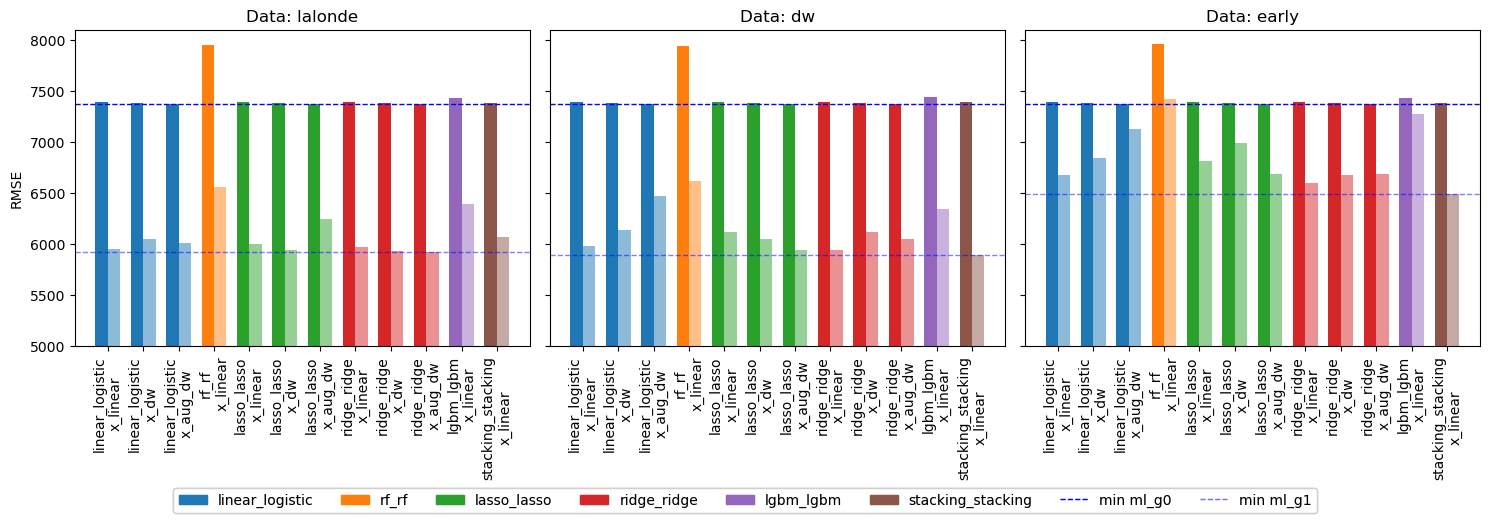}
    \caption{Root mean squared error for different learners and model specifications for the nuisance components $g(0,X)$ and $g(1,X)$ evaluated on the out-of-sample predictions from cross-fitting. LaLonde data example.}
    \label{fig:rmse_lalonde}
\end{figure}

\begin{figure}
    \centering
    \includegraphics[width=0.85\linewidth]{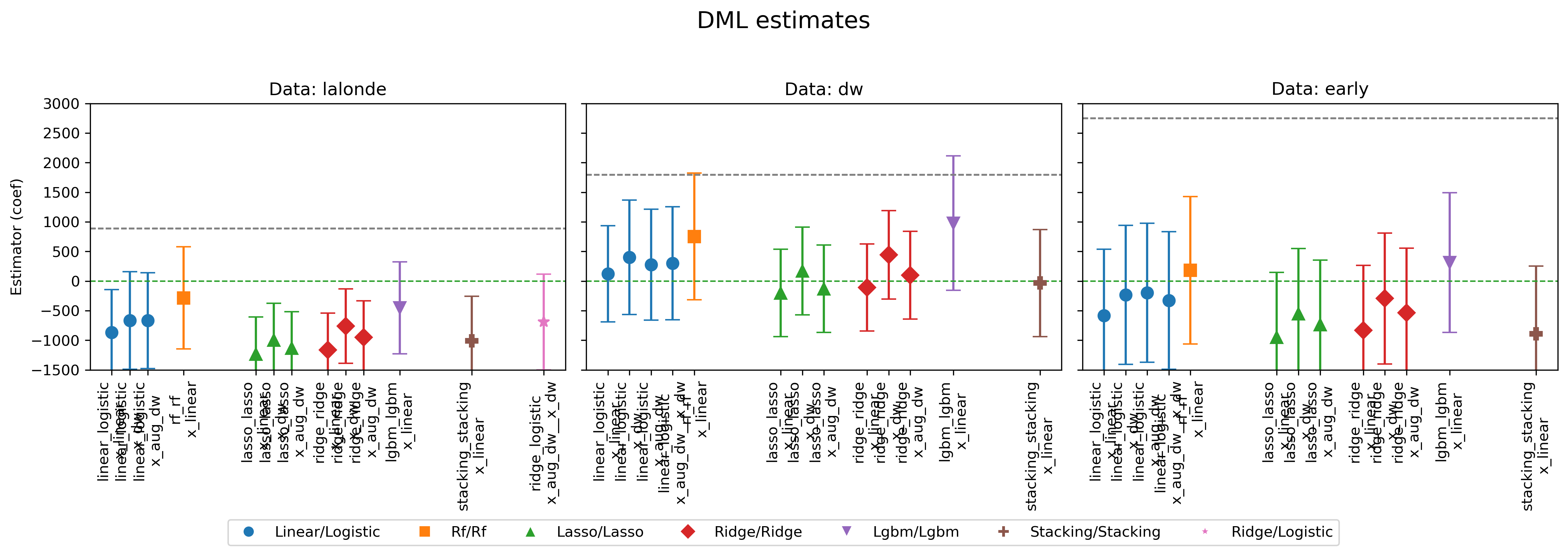}
    \caption{DML DiD estimates corresponding to the different learner and model specifications, Lalonde data example.}
    \label{fig:dml_lalonde}
\end{figure}

The empirical application presented in Section \ref{app:lalonde} is an extended replication of the analysis in \cite{sant2020doubly}, who evaluate the performance of the doubly robust ATT estimator in the canonical DiD setting based on the data and previous results by \cite{smith2005}. We use exactly the same data files from the replication materials of \cite{sant2020doubly} for the ``LaLonde``, ``DW`` and ``early randomized'' data sets considered in \cite{smith2005}. We extend the previous results by reporting machine-learning based estimates of the ATT using unpenalized and penalized linear and logistic regression regression in three different specifications: Linear in covariates, linear in the specification of DW (using polynomials on age, education, and income) and an augmented specification including interaction of some covariates, more details provided in Table \ref{tab:covariates} and \cite{sant2020doubly}. For penalized estimation in the basic and augmented specifications, we use cross-validated lasso and ridge estimation. Moreover, we evaluate the performance of non-linear machine learners including random forests, gradient boosting, and a stacking ensemble learner combining the previously mentioned algorithms. We use scikit-learn's robust scaler for all learners and calibrate propensity scores with isotonic regression. The results in the main text refer to the ``best'' specification choosing the learners with the lowest prediction error (root mean squared error for regression prediction and log loss for propensity score). Results of alternative parametrizations and first-stage prediction errors are shown in Figures \ref{fig:log_loss_lalonde} and \ref{fig:rmse_lalonde} with corresponding DML estimates presented in Figure \ref{fig:dml_lalonde}.

\begin{table}[htbp]
\centering
\begin{tabular}{|p{4cm}|p{10cm}|}
\hline
\textbf{Specification} & \textbf{Variables} \\
\hline
Linear & age, educ, black, married, nodegree, hisp, re74 \\
\hline
DW & age, educ, black, married, nodegree, hisp, re74, ze74, agesq, agecub, educsq, educre74 \\
\hline
Augmented DW & age, educ, black, married, nodegree, hisp, re74, ze74, agesq, agecub, educsq, educre74, mare74, maze74 \\
\hline
\end{tabular}
\caption{Covariate Specifications. More details available in \cite{sant2020doubly}.}
\label{tab:covariates}
\end{table}

\newpage

\subsection{Estimating the Effect of Minimum Wages on Firm Profitability}

\begin{table}[ht]
\centering
\begin{tabular}{lrrrrrr}
\toprule
 & coef & std err & t & P>|t| & 2.5 \% & 97.5 \% \\
\midrule
ATT(2000,1997.0,1998.0) & -0.001 & 0.008 & -0.097 & 0.923 & -0.017 & 0.016 \\
ATT(2000,1998.0,1999.0) & 0.007 & 0.010 & 0.648 & 0.517 & -0.014 & 0.027 \\
ATT(2000,1999.0,2000.0) & -0.021 & 0.010 & -2.068 & 0.039 & -0.041 & -0.001 \\
ATT(2000,1999.0,2001.0) & 0.031 & 0.019 & 1.620 & 0.105 & -0.007 & 0.069 \\
ATT(2000,1999.0,2002.0) & 0.020 & 0.019 & 1.019 & 0.308 & -0.018 & 0.058 \\
\bottomrule
\end{tabular}
\caption{Output from DML point estimation for $ATT(\g,t)$ parameters in the replication of \cite{draca2011minimum}. Dependent variable is net profit margin.}
\label{tab:attgt_draca}
\end{table}

In the following, we present additional results in the empirical application based on \cite{draca2011minimum} evaluating the effect of the NWM on log average wages as the outcome variable (\texttt{ln\_avwage}). Table \ref{tab:attgt_draca} presents the detailed estimation results for the empirical application in Section \ref{sec:draca}.

\begin{figure}[t]
    \centering
    \includegraphics[width=\linewidth]{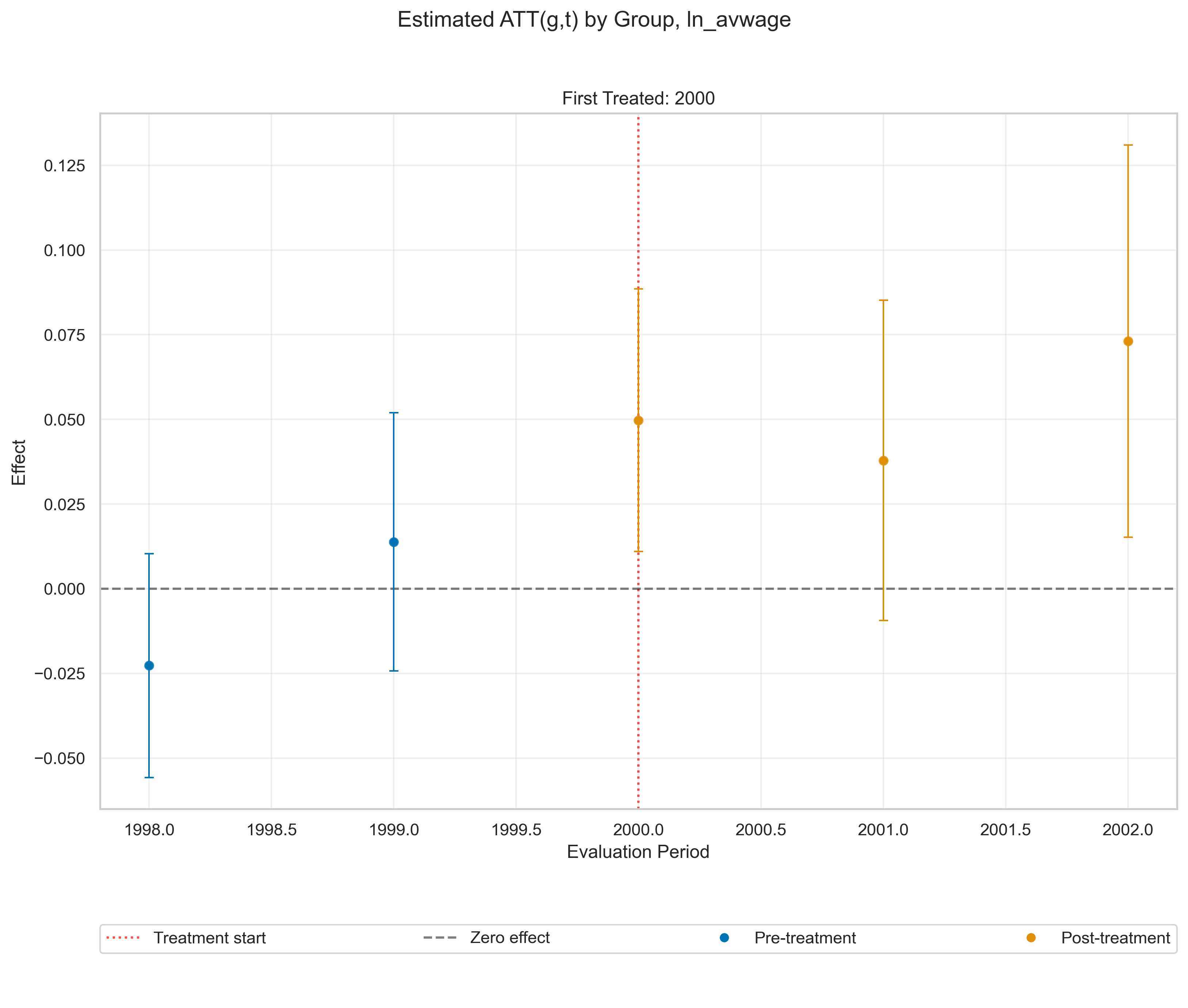} 
    \caption{$ATT(\g,t)$ estimates and $95\%$ (pointwise) confidence intervals associated with the 1999-2000 NWM introduction on average wages. Dependent variable \texttt{ln\_avwage}. Reanalysis of \cite{draca2011minimum}.}
    \label{fig:attgt_draca_wages}
\end{figure}

\begin{figure}[ht]
    \centering
    \includegraphics[width=0.7\linewidth]{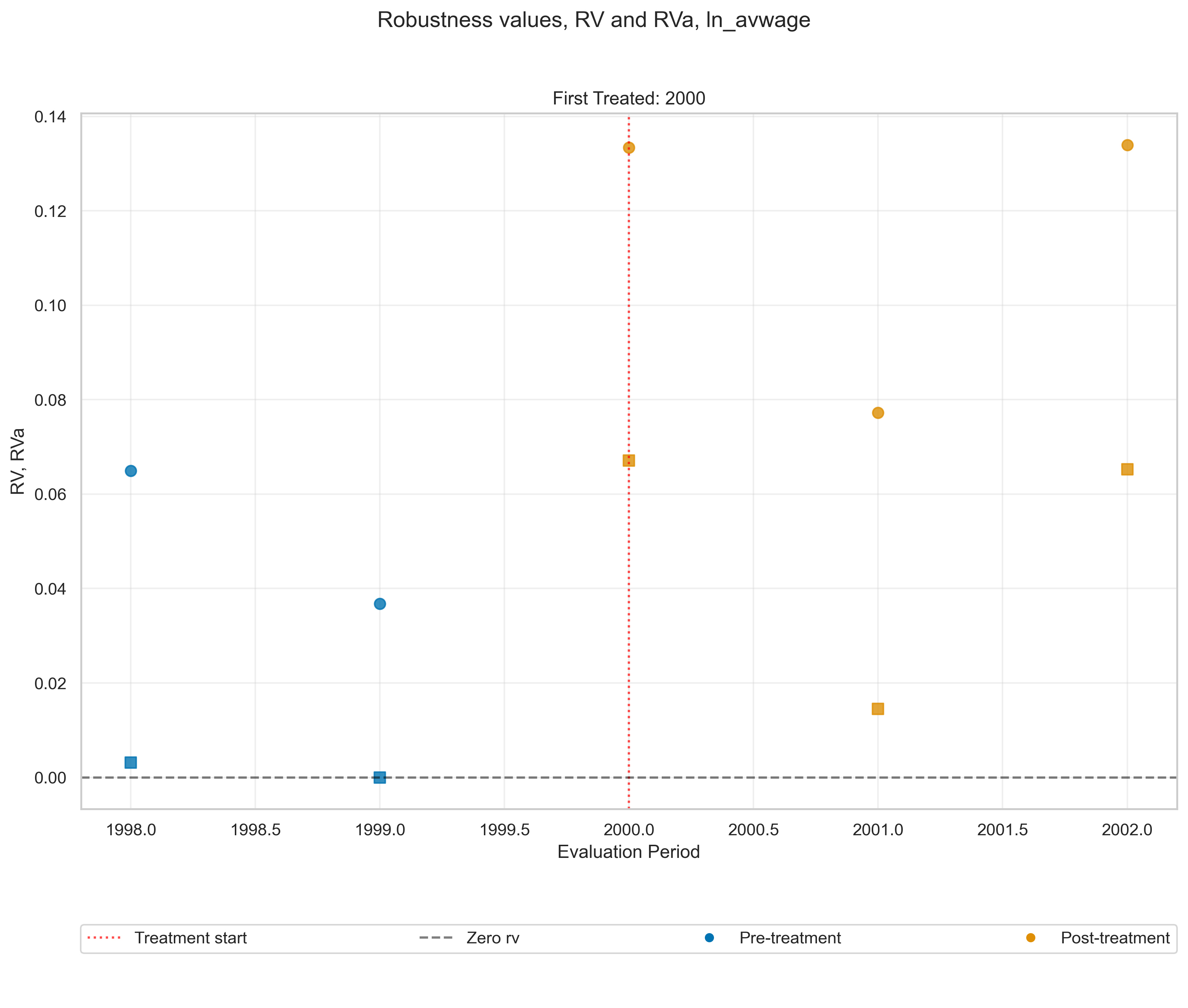} 
    \caption{Robustness values, $RV$ and $RV_{a=0.1}$ for $ATT(\g,t)$ parameters for the 1999-2000 NWM introduction on average wages. Dependent variable \texttt{ln\_avwage}. Reanalysis of \cite{draca2011minimum}.}
    \label{fig:rvs_draca_wages}
\end{figure}

\begin{table}[t]
    \centering
\begin{tabular}{l lrrr}
\toprule
 Dependent variable & Scenario & cf\_d & cf\_y & abs(rho) \\
\midrule
ln\_avwage & \textit{Pretest} & 0.0650 & 0.0650 & 1.0000 \\
& \textit{Benchmark}  & & & \\
& - ptwk &  0.0972 &  0.0010 &  1.0000  \\
& - sic2 & 0.1429 & 0.0010 & 1.0000 \\
& - unionmen & 0.0620 & 0.0010 & 1.0000 \\
& - female & 0.0309 & 0.0010 & 1.0000\\
& - gorwk & 0.0416 & 0.0010 & 1.0000  \\\\
\bottomrule
\end{tabular}
    \caption{PT violation scenarios based on pre-testing and benchmarking pre-treatment covariates. We enforce a minimum value of $0.0010$ for benchmarking scenarios. Dependent variable \texttt{ln\_avwage}. Reanalysis of \cite{draca2011minimum}.}
    \label{tab:bench_wages}
\end{table}

\begin{figure}[ht]
    \centering
    \includegraphics[width=0.7\linewidth]{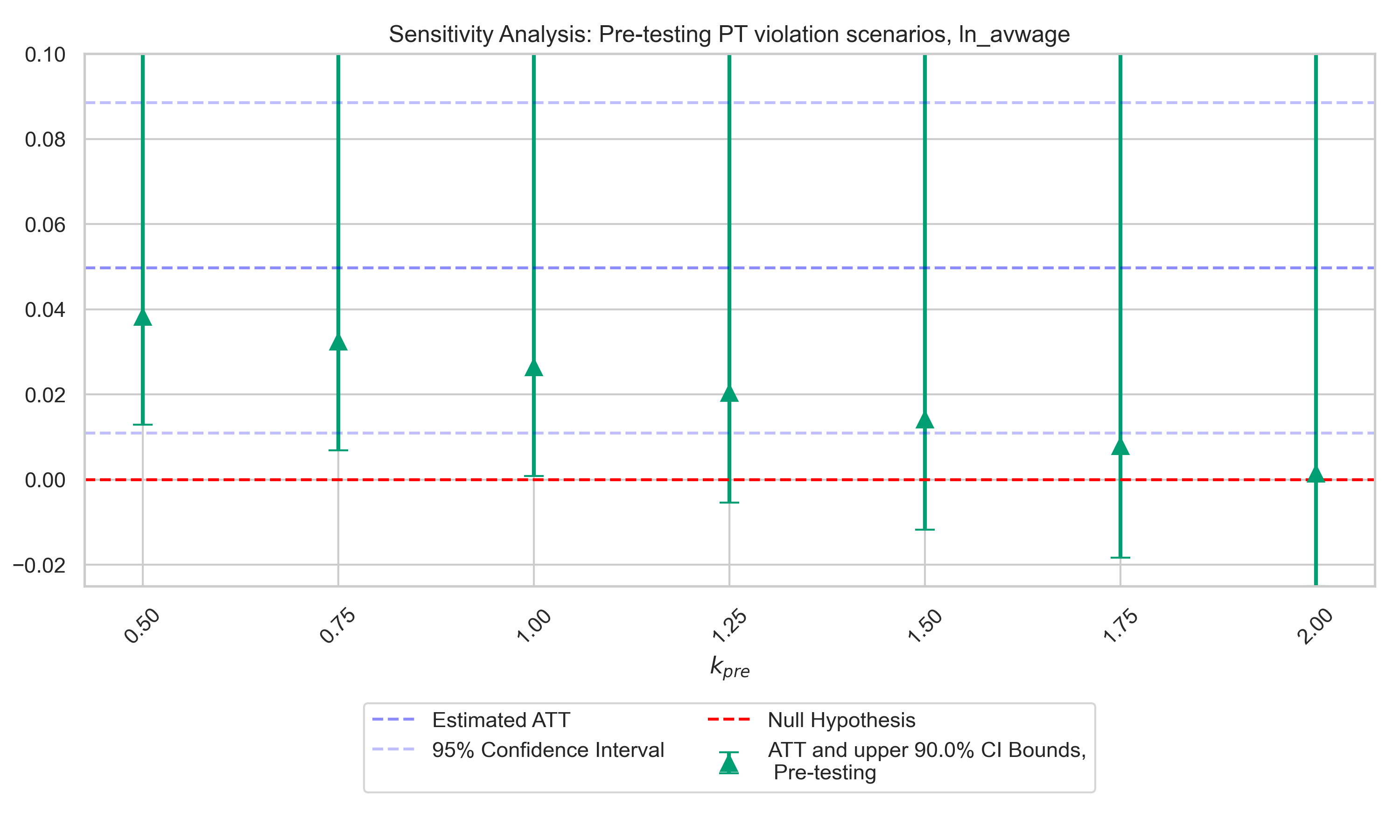}
    \caption{Lower sensitivity bound and one-sided $90\%$ confidence sensitivity bounds according for the $ATT(\g,t)$ in the first post-treatment period in different parallel trend violation scenarios based on pre-testing for dependent variable \textit{ln\_avwage}. Reanalysis of \cite{draca2011minimum}.}
    \label{fig:bounds_draca_wage}
\end{figure}

\end{document}